\documentclass[pra,onecolumn,floatfix,a4paper,superscriptaddress]{revtex4}
\usepackage{bm,color,graphicx,amsmath,txfonts}
\usepackage{here}
\usepackage{array}
\usepackage{graphicx}
\usepackage[colorlinks, citecolor=blue,linkcolor=blue]{hyperref}
\usepackage{braket}
\usepackage{dsfont}
\begin{document}

\makeatletter
\renewcommand{\@biblabel}[1]{\makebox[2em][l]{\textsuperscript{\textcolor{black}{\fontsize{10}{12}\selectfont[#1]}}}}
\makeatother

\let\oldbibliography\thebibliography
\renewcommand{\thebibliography}[1]{%
  \addcontentsline{toc}{section}{\refname}%
  \oldbibliography{#1}%
  \setlength\itemsep{0pt}%
}

\title{Probing multiparameter quantum estimation in the process $e^+e^-\to J/\psi \to \text{B}\bar{\text{B}}$ at BESIII}

\author{Elhabib Jaloum}
\affiliation{LPTHE-Department of Physics, Faculty of Sciences, Ibnou Zohr University, Agadir, Morocco}

\author{Mohamed Amazioug}\email{m.amazioug@uiz.ac.ma}
\affiliation{LPTHE-Department of Physics, Faculty of Sciences, Ibnou Zohr University, Agadir, Morocco}

\date{\today}

\begin{abstract}

The quantum Fisher information matrix (QFIM) is the cornerstone of multiparameter quantum metrology. In this work, we investigate multiparameter quantum estimation in baryon-antibaryon ($\text{B}\bar{\text{B}}$) pairs produced via the $e^+e^-\to J/\psi \to \text{B}\bar{\text{B}}$ process at the BESIII experiment, utilizing the symmetric logarithmic derivative (SLD) formalism. Moreover, the QFIM defines the quantum Cram\'er-Rao bound and dictates the choice of optimal probe states. We compare individual and simultaneous estimation strategies for two key physical parameters: the scattering angle $\varphi$ and the decay parameter $\alpha_{\psi}$. The estimation variances are found to depend strongly on the explored region of the $(\varphi, \alpha_{\psi})$ parameter space and to display markedly different temporal dynamics. In general, higher true values of a parameter increase the system’s sensitivity, thereby significantly reducing the associated variance. While both variances increase with evolution time, they do so at distinct rates, revealing parameter-dependent information loss driven by environmental decoherence. These findings demonstrate the utility of the QFIM framework for multiparameter quantum estimation in realistic open systems and provide new insights into the ultimate precision limits achievable for hyperon decay parameters.

\end{abstract}
\maketitle

\section{Introduction}    \label{sec:1}

Parameter estimation plays a central role in the development of ultra-high-precision technologies across multiple fields \cite{ref8,ref10}. Moreover, Fiber-optic Mach-Zehnder (MZI) and Sagnac interferometers are fundamental architectures for physical parameter estimation—a central challenge governed by fundamental limits such as the Cram\'er–Rao bound. The MZI, in particular, is a cornerstone device for phase estimation \citep{Int6} and find widespread applications in precision measurement \citep{Int1}, quantum communication \citep{Int2}, quantum sensing \citep{Int3}, state estimation \citep{Int4}, and quantum machine learning \citep{Int5}. Similarly, the Sagnac interferometer enables the simultaneous estimation of multiple quantities (multiparameter estimation), including rotation via fiber-optic gyroscopes (FOGs) \citep{Int7}, current \citep{Int8}, temperature \citep{Int9}, and strain \citep{Int10}, among others \citep{Int11}. In recent years, quantum metrology has drawn significant interest by exploiting genuinely quantum features to surpass the precision limits imposed by classical strategies and to enable novel measurement protocols \cite{ref11,ref12}. Prominent applications of quantum metrology now include clock synchronization \cite{ref13}, enhancement of gravitational-wave detector sensitivity \cite{ref14}, derivation of ultimate bounds on phase estimation \cite{ref15,ref16}, estimation of space–time parameters \cite{ref18,ref19,ref20}, electromagnetic field sensing \cite{ref21,ref22}, and optimal thermometry of quantum reservoirs \cite{ref23,ref24,ref25}. These quantum metrology protocols achieve substantial precision gains by harnessing quantum correlations in multipartite systems, notably entanglement \cite{ref26,ref27,ref28} and quantum discord \cite{ref29,ref30}. \par

The ultimate precision limit for estimating a set of parameters $\hat{\theta}$ in quantum metrology is given by the Quantum Cram\'er-Rao Bound \cite{ref31}, which states that the covariance matrix satisfies 
\begin{equation}{\rm Cov}(\hat{\theta}) \ge F^{-1},
\end{equation}
where $\mathcal{F}$ is the Quantum Fisher Information Matrix (QFIM) \cite{ref32}. For the estimation of a single parameter $\theta$, this inequality simplifies to the scalar form 
\begin{equation}
{\rm Var}(\theta) \ge \mathcal{F}^{-1},
\end{equation}
with $\mathcal{F}$ being the Quantum Fisher Information (QFI) \cite{ref33,ref35}. Lower variance corresponds to higher precision, so the primary objective of any quantum metrology protocol is to minimize this variance-or equivalently, to maximize the QFI (or the QFIM in the multiparameter case). The inverse of the QFIM thus provides the fundamental lower bound on estimation error. Single-parameter estimation has been thoroughly investigated \cite{ref36,ref37}, largely because an optimal probe state that maximizes the QFI typically exists \cite{ref38,ref39}. In contrast, realistic scenarios generally involve multiple unknown parameters, for which no universally optimal probe state exists that simultaneously maximizes all elements of the QFIM \cite{ref40}. Moreover, the Quantum Cram\'er-Rao Bound is not always attainable due to the potential incompatibility of optimal measurements for different parameters \cite{ref41,ref42}. Consequently, simultaneous multiparameter estimation has emerged as a critical and challenging topic in quantum metrology. Recent works have demonstrated that joint estimation of multiple parameters can yield higher precision than separate individual estimations. In particular, entanglement among multiple particles has been shown to enhance multiphase estimation \cite{ref43,ref44}, while two-mode entangled coherent states have been proposed for the simultaneous estimation of linear and nonlinear phase shifts \cite{ref45}. \par

The improvement of precision in quantum metrology has been attributed in several studies to the presence of quantum correlations, recognized as an essential resource \cite{ref46}. In this regard, studies have addressed the potential of the QFIM (or QFI) enhancement to not only signal the presence of quantum correlations within a multipartite system but also to establish the Quantum Fisher Information as a quantifier for these correlations. Works have been developed to address this connection \cite{ref47,ref48}. Crucially, however, the comprehensive understanding of how quantum correlations and entanglement fundamentally influence and enable the highest precision in quantum metrology is far from complete. \par

Quantum entanglement has attracted growing interest in high-energy physics, notably through the study of hadronic final states resulting from proton collisions \cite{reff26,reff27}, enabling exploration at smaller length scales \cite{ref28}. Systems such as neutral kaons \cite{reff29,reff30,reff31,reff32}, neutral B mesons \cite{reff33}, and positronium \cite{ref34} are particularly promising for testing entanglement and CP violation. Recently, studies have highlighted entanglement in top-quark pair production at the LHC \cite{reff35} and the experimental feasibility of Bell inequality violation \cite{reff36}. This has motivated further investigations into entanglement in top-quark production \cite{reff37,reff38,reff39,reff40,reff41}, hyperons \cite{reff42}, as well as gauge bosons from Higgs boson decay or direct production \cite{reff43,reff44,reff45,reff46}. Entanglement has been observed at $\sqrt{s} = 13\ \text{TeV}$ \cite{reff47}, and Bell inequality violation has been suggested in B meson decays at the LHCb and Belle II experiments \cite{reff48}. \par

Polarization correlations have mainly been studied for baryons with spin $1/2$ or $3/2$ and well-established parity \cite{reff57,reff58,reff59}, notably by the BESIII Collaboration for ground-state octet and decuplet baryons \cite{reff66,reff67}. For excited baryons, determining both parity and whether the spin exceeds $3/2$ remains essential. Spin density matrices are formulated in two main ways: the standard form \cite{reff68,reff69,reff70,reff71}, which is suitable for high spins but less intuitive, and the Cartesian form \cite{reff72,reff73,reff74,reff75,reff76}, which offers a clearer physical interpretation but lacks a general decomposition method for high-spin cases. Although both forms are mathematically consistent, no direct mapping between them has yet been established. \par

The study of effective transition form factors is essential for understanding baryon production in processes such as electron-positron annihilation, where various baryons can be produced via virtual photons or resonance states \cite{reff77,reff78,reff79,reff80,reff81,reff82,reff83}. These transition form factors are described within the helicity formalism, which captures the spin and polarization properties of the particles, allowing the calculation of helicity transition amplitudes and polarization correlation coefficients in two-baryon systems \cite{reff84}. This framework not only facilitates the analysis of baryon properties but also enables the exploration of CP violation by comparing the production processes of baryons and their antiparticles, thereby highlighting the matter-antimatter asymmetry. The interplay of these concepts thus provides a comprehensive approach to studying baryon dynamics and fundamental symmetries in particle physics.

In this work, we investigate a multiparameter quantum estimation strategy in the context of quantum metrology using SLD in the process $e^+e^- \to J/\psi \to \text{B} \bar{\text{B}}$, where $\text{B}$ and $\bar{\text{B}}$ denote a spin-$1/2$ baryon and its antibaryon, respectively, using the BESIII experiment. We derive the multiparameter Quantum Cram\'er-Rao Bound for the simultaneous and individual estimation of the scattering angle $\varphi$ and the decay parameter of the vector charmonium state $\psi(c\bar{c})$ (typically the $J/\psi$). This bound is obtained from the QFIM. Furthermore, we compare the performance of individual and joint estimation strategies and demonstrate that the highest precision for both parameters is systematically achieved through the simultaneous measurement approach.

The article is structured as follows: Section~\ref{sec:2} recalls the fundamental principles of multiparameter quantum estimation, along with the mathematical tools required to derive the quantum Fisher information matrix (QFIM), with particular emphasis on the vectorization technique applied to the density matrix. Section~\ref{sec:3} investigates the ultimate precision attainable in multiparameter estimation in the stationary regime, using the QFIM, for the process \( e^+ e^- \to J/\psi\to \text{B}\bar{\text{B}} \). Section~\ref{sec:4} is dedicated to the precise estimation of the scattering angle $\varphi$ and the decay parameter $\alpha_{\psi}$ associated with the vector charmonium $\psi(c\bar{c})$. Section~\ref{sec:5} subsequently analyzes the dynamical behavior of simultaneous and individual estimation strategies in both Markovian and non-Markovian regimes. Finally, the main conclusions are presented in Section~\ref{sec:6}.

\section{Quantum Fisher information matrix (QFIM)}\label{sec:2}

This section is dedicated to reviewing the mathematical framework necessary for the derivation of the QFIM. We avoid the need for matrix diagonalization by converting the density matrix into a column vector, allowing for a more direct algebraic evaluation of the QFIM. By considering the Hilbert-Schmidt space $\mathbb{K}^{n \times n}$, the vec-operator is introduced to map any matrix $\mathbb{X}$ into a column vector according to \cite{Gilchrist,Dd}

\begin{equation}
{\rm vec}\left[ \mathbb{X} \right] = {\left( {{\pi_{11}},...,{\pi_{n1}},{\pi_{12}},...,{\pi_{n2}},...,{\pi_{1n}},...,{\pi_{nn}}} \right)^T}.
\end{equation}
Furthermore, using the expression $\mathbb{X} = \sum\limits_{\alpha,\beta=1}^n {{a_{\alpha\beta}}} \left| \alpha \right\rangle \left\langle \beta \right|$, the action of the vec-operator on a matrix $\mathbb{X}$ is written as
\begin{equation}
{\rm vec}\left[\mathbb{X} \right] = \left( {{\mathbb{I}_{n \times n}} \otimes \mathbb{X}} \right)\sum\limits_{i=1}^n {{b_i} \otimes {b_i}}, \label{vecA}
\end{equation}
where ${b_i}$ represent the elements of the computational basis for the space ${\mathbb{K}^{n \times n}}$. The result of applying the vec-operator to matrix $\mathbb{X}$ is a single column vector whose components are the columns of $\mathbb{X}$ ordered from first to last. Exploiting the properties of the Kronecker product \cite{Schacke2004}, one can derive
\begin{equation}
{\rm vec}\left[ {\mathbb{XY}} \right] = \left( {{\mathbb{I}_n} \otimes \mathbb{X}} \right){\rm vec}\left[ \mathbb{Y} \right] = \left( {{\mathbb{Y}^T} \otimes {\mathbb{I}_n}} \right){\rm vec}\left[ \mathbb{X} \right], \label{vec1}
\end{equation}
\begin{equation}
{\rm tr}\left( {{\mathbb{X}^\dag }\mathbb{Y}} \right) = {\rm vec}{\left[\mathbb{X} \right]^\dag }{\rm vec}\left[ \mathbb{Y}\right]. \label{vec3}
\end{equation}
\begin{equation}
{\rm vec}\left[ {\mathbb{XZY}} \right] = \left( {{\mathbb{Y}^T} \otimes \mathbb{X}} \right){\rm vec}\left[ \mathbb{Z} \right],\label{vec2}
\end{equation}
After establishing these matrix relationships, we transition to the derivation of the QFIM using the vectorized density matrix $\text{vec}(\varrho)$. For clarity, we begin by reviewing the general definition of the QFIM for the simultaneous estimation of the parameters $\theta_1, \dots, \theta_n$. The QFI represents the ultimate information limit obtainable for an estimated parameter $\theta$ from a given state $\varrho_\theta$. For a single-parameter state, the QFI expression is given by
$
\mathcal{F}\left( {{\varrho _\theta }} \right) = {\rm Tr}\left\{ {{\varrho _\theta }{L_\theta }^2} \right\}
$
where $L_\theta$ is the associated Symmetric Logarithmic Derivative (SLD). For multiparameter estimation tasks, the ultimate precision limits for the set $\{\theta_i\}$ are determined by the QFI \cite{ref31,Eg}.
\begin{equation}
{F_{ij}} = \frac{1}{2}{\rm Tr}\left\{ {\left( {{{\hat L}_{\theta _i}}{{\hat L}_{\theta _j}} + {{\hat L}_{\theta _j}}{{\hat L}_{\theta _i}}} \right) \varrho } \right\}, \label{F}
\end{equation}
The SLDs, denoted ${{{\hat L}_{\theta _i}}}$, are defined by the following algebraic equations
\begin{equation}
{\partial _{\theta _i}} \varrho  = ({\hat L_{\theta _i}} \varrho  + \hat \varrho {\hat L_{\theta _i}})/2. \label{L}
\end{equation}
To arrive at the explicit form of the QFIM (Eq. (\ref{F})), one must first solve for the SLD ${{{\hat L}_{\theta _i}}}$ using the algebraic relationship provided in Eq. (\ref{L}). Several explicit expressions for the QFIM have been reported \cite{Banchi2014,Sommers2003,Paris2009}. Notably, the QFIM was derived in terms of the eigenvalues of the density matrix $\varrho$ by employing its spectral decomposition,
$
{\varrho } = \sum\limits_\alpha {{p_\alpha}} \left| \alpha \right\rangle \left\langle \alpha \right|
$
\cite{Banchi2014,Sommers2003}.
\begin{equation}
    {F_{ij}} = 2\sum\limits_{{p_\alpha} + {p_\beta} > 0} {\frac{{\left\langle \alpha \right|{\partial _{{\theta _i}}}{\varrho }\left| \beta \right\rangle \left\langle \beta \right|{\partial _{{\theta _j}}}{\varrho }\left| \alpha \right\rangle }}{{{p_\alpha} + {p_\beta}}}}, \label{F1}
\end{equation}
consequently, the SLDs are expressed as
\begin{equation}
    {L_{{\theta _i}}} = 2\sum\limits_{{p_\alpha} + {p_\beta} > 0} {\frac{{\left\langle \alpha \right|{\partial _{{\theta _i}}}{\varrho }\left| \beta \right\rangle }}{{{p_\alpha} + {p_\beta}}}} \left| \alpha \right\rangle \left\langle \beta \right|.
\end{equation}
In accordance with the results of Paris \cite{Paris2009}, the QFIM is derived through a representation of the density matrix that facilitates the computation of the symmetric logarithmic derivatives
\begin{equation}
    {F_{ij}} = 2\int\limits_0^\infty  {{\rm Tr}\left[ {{e^{ - {\varrho }t}}{\partial _{{\theta _i}}}{\varrho }{e^{ - {\varrho }t}}{\partial _{{\theta _j}}}{\varrho }} \right]}. \label{F2}
\end{equation}
\v{S}afr\'{a}nek \cite{Safranek2018} recently proposed a distinct explicit expression for the QFIM based on the vectorization of the density matrix $\varrho$. This method is particularly advantageous because it is analytically applicable to systems of arbitrary dimension. Furthermore, it avoids the diagonalization required by Eq. \ref{F1} and removes the integral and exponentiation requirements of Eq. \ref{F2}. The calculation involves inverting the following matrix
\begin{equation}
\Lambda  = \left( {{\varrho ^T} \otimes \mathbb{I} + \mathbb{I} \otimes \varrho } \right). \label{landa}
\end{equation}
The QFIM (Eqs. \ref{F1} and \ref{F2}) can be transformed into the following expression using the properties given by Eqs. \ref{vec1}, \ref{vec2}, and \ref{vec3}
\begin{equation}
{F_{ij}} = 2{\rm vec}{\left[ {{\partial _i}\hat \varrho } \right]^T}{\Lambda ^{+}}{\rm vec}\left[ {{\partial _j}\hat \varrho } \right], \label{Flanda}
\end{equation}
and the SLDs are written as
\begin{equation}
{\rm vec}\left[ {L_{{\theta _i}}} \right] = 2{\Lambda ^{+}}{\rm vec}\left[ {{\partial _i}\hat \varrho } \right]. \label{vecL}
\end{equation}
The saturation of the scalar Cram\'er-Rao bound,
$
{\rm Var}\left( \theta \right) \ge {\mathcal{F}^{ - 1}}
$
is always achievable in single-parameter scenarios. This saturation defines the optimal quantum measurement, which utilizes the projectors corresponding to the eigenvectors of the SLD ${L_\theta }$. In contrast to single-parameter estimation, the saturability of the matrix Cram\'er-Rao inequality,
$
{\rm Cov}\left( {\hat \theta } \right) \ge {F^{ - 1}}
$
is not guaranteed in the multiparameter setting. This is fundamentally due to the potential incompatibility (non-commutativity) of the optimal measurement operators for the different parameters \cite{Aj1,Rehacek2018}. It is therefore necessary to investigate the conditions required to ensure the saturation of the bound. This investigation involves first solving Eq. \ref{vecL} for the SLDs $L_{{\theta _i}}$. If these operators, $L_{{\theta _i}}$, are found to commute, then a common eigenbasis for all SLDs can be established. The existence of a common eigenbasis allows for a simultaneous measurement that achieves the optimal precision defined by the Cram\'er-Rao inequality. While a straightforward way to achieve this is via the commutativity of the SLDs ($\left[ {L_{{\theta _i}},L_{{\theta _j}}} \right] = 0$), this condition is sufficient but not necessary. For the case where the SLDs are non-commuting, the Cram\'er-Rao inequality is saturated if and only if the weaker condition,
$
\text{Tr}\left( {\varrho \left[ {L_{{\theta _i}},L_{{\theta _j}}} \right]} \right) = 0,
$
is satisfied \cite{Matsumoto2002,Crowley2014}.

\section{Theoretical model}\label{sec:3}

\begin{figure}[!h]
\begin{center}
\includegraphics[width=8.5cm,height=6cm]{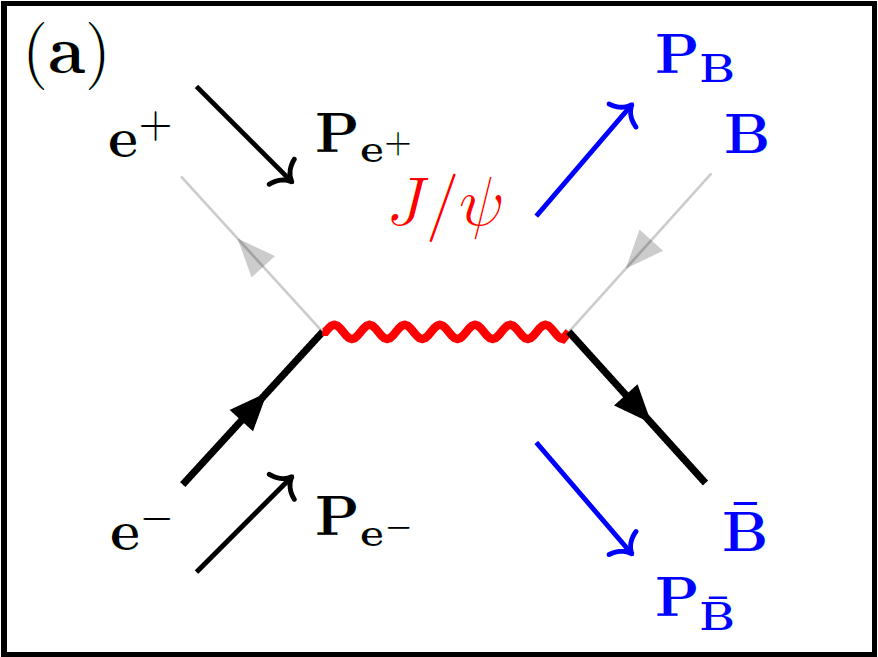}
\includegraphics[width=8.5cm,height=6cm]{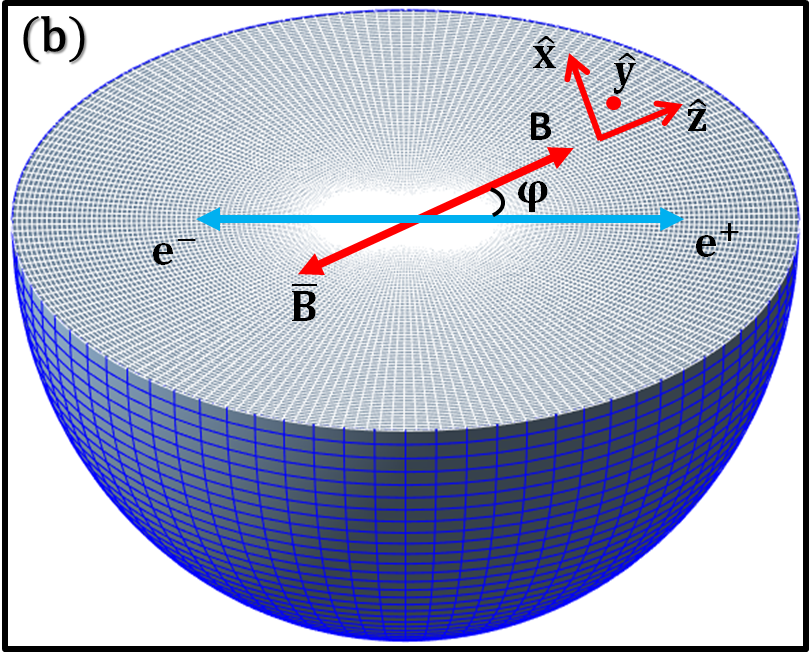}
\end{center}
\caption{(a) The Feynman diagram illustrating the annihilation process $e^{+}e^{-}\to \text{B}\bar{\text{B}}$. (b) The orientation of the coordinate axes $\{\hat{x}, \hat{y}, \hat{z}\}$, established in the mutual rest frame of the $\text{B}$ and $\bar{\text{B}}$ pair.}
\label{fig:e}
\end{figure}

For a pair of spin-$1/2$ particles—exemplified here by the $\text{B}\bar{\text{B}}$ system—the general form of the quantum state is represented by the density matrix
\begin{equation}
\varrho_{\text{B}\bar{\text{B}}} = \frac{1}{4}\sum_{\mu,\nu=0}^{3}\mathcal{S}_{\mu\nu}\Big(\tau^{\text{B}}_{\mu} \otimes \tau^{\bar{\text{B}}}_{\nu}\Big),
\label{eq:BB}
\end{equation}
we utilize the Pauli matrices $\tau^{\text{B}}_{\mu}$ and $\tau^{\bar{\text{B}}}_{\nu}$ to span the spin spaces of the baryon and antibaryon in their center-of-mass frame. Polarizations and spin correlations are quantified by the $4\times4$ real matrix $\mathcal{S}_{\mu\nu}$, with a primary focus on $\text{B}\bar{\text{B}}$ pairs produced through the $e^+e^- \to J/\psi$ production chain. We define the spin matrices $\tau^{\text{B}}_{\mu}$ and $\tau^{\bar{\text{B}}}_{\nu}$ within the respective reference frames of the baryon and antibaryon, denoted by the basis vectors $(\mathbf{\hat{x}}_{\text{B}},\mathbf{\hat{y}}_{\text{B}},\mathbf{\hat{z}}_{\text{B}})$ and $(\mathbf{\hat{x}}_{\bar{\text{B}}},\mathbf{\hat{y}}_{\bar{\text{B}}},\mathbf{\hat{z}}_{\bar{\text{B}}})$. Similarly, the rest frame of the antibaryon $\bar{\text{B}}$ is defined using the same axes as the baryon, $(\mathbf{\hat{x}}_{\bar{\text{B}}},\mathbf{\hat{y}}_{\bar{\text{B}}},\mathbf{\hat{z}}_{\bar{\text{B}}}) = (\mathbf{\hat{x}},\mathbf{\hat{y}},\mathbf{\hat{z}})$, as depicted in Fig. \ref{fig:e}(a).
To the leading order, the spin-correlation matrix $\mathcal{S}_{\mu\nu}$ for the $\text{e}^{+}\text{e}^{-} \to J/\psi \to \text{B}\bar{\text{B}}$ process is parameterized by the decay asymmetry $\alpha_{\psi} \in [-1, 1]$ and the relative phase $\Delta\Phi \in [-\pi, \pi]$. The elements of $\mathcal{S}_{\mu\nu}$ are expressed as functions of the baryon production angle $\varphi$ \cite{Aj2,Aj3,Aj4,JLAM}
\begin{equation}
\mathcal{S}_{\mu\nu} =
\begin{pmatrix}
\mathcal{S}_{0,0} & 0 & \mathcal{S}_{0,y} & 0 \\
0 & \mathcal{S}_{x,x}  & 0 & \mathcal{S}_{x,z}  \\
\mathcal{S}_{y,0} & 0 & \mathcal{S}_{y,y} & 0 \\
0 & \mathcal{S}_{z,x}  & 0 & \mathcal{S}_{z,z} 
\end{pmatrix}, \quad \text{where} \quad
\begin{aligned}
\mathcal{S}_{0,0} &= 1, \quad \mathcal{S}_{0,y}= \mathcal{S}_{y,0} = \frac{ \beta_{\psi}\cos\varphi\sin\varphi}{1+\alpha_{\psi}\cos^{2}\varphi}, \\
\mathcal{S}_{x,x}&=\frac{\sin^{2}\varphi}{1+\alpha_{\psi}\cos^{2}\varphi}, \quad \mathcal{S}_{x,z}=\mathcal{S}_{x,z} =\frac{\gamma_{\psi}\cos\varphi\sin\varphi}{1+\alpha_{\psi}\cos^{2}\varphi},\\
\mathcal{S}_{z,z}&=\frac{-\alpha_{\psi}\sin^{2}\varphi}{1+\alpha_{\psi}\cos^{2}\varphi}, \quad \mathcal{S}_{y,y}=\frac{\alpha_{\psi}+\cos^{2}\varphi}{1+\alpha_{\psi}\cos^{2}\varphi},
\end{aligned}
\label{M:1}
\end{equation}
the variables $\beta_\psi$ and $\gamma_\psi$ are given as functions of $\alpha_\psi$ and $\Delta\Phi$ by primary parameters $\alpha_\psi$ and $\Delta\Phi$ as
\begin{subequations}
\begin{align}
\beta_\psi = \sqrt{1 - \alpha_\psi^2} \sin(\Delta\Phi),
\end{align}
\begin{align} \quad \gamma_\psi = \sqrt{1 - \alpha_\psi^2} \cos(\Delta\Phi),
\end{align}
and
\begin{align} \quad \alpha_{\psi}+\beta_{\psi}+\gamma_{\psi}=1.
\end{align}
\end{subequations}
The normalization constant in Eq. (\ref{M:1}) is determined by this identity. Relative to the baryon's local rest frame $(\mathbf{\hat{x}}, \mathbf{\hat{y}}, \mathbf{\hat{z}})$, the polarization vector $\mathbf{P}_{\text{B}}$ is expressed as
\begin{equation}
\mathbf{P}_{\text{B}}=\frac{\beta_{\psi}\cos(\varphi)\sin(\varphi)}{1+\alpha_{\psi}\cos(\varphi)}\mathbf{\hat{y}}_{1},
\end{equation}
the hyperon and antihyperon polarization vectors are equivalent, such that $\mathbf{P}_{\text{B}} = \mathbf{P}_{\bar{\text{B}}}$. To facilitate the analysis, we apply a coordinate transformation $(\mathbf{\hat{y}} \leftrightarrow \mathbf{\hat{z}})$ to cast the state into an X-structure. The transformed spin density matrix is expressed as
\begin{equation}
\varrho_{\text{Y}\bar{\text{Y}}}^{\text{X}} = \frac{1}{4} \left( \mathbb{I}\otimes \mathbb{I} + \mathbb{A}_{z,0}\big(\tau_z \otimes \mathbb{I}\big) + \mathbb{A}_{0,z}\big(\mathbb{I} \otimes \tau_z\big) + \sum_{i=j} \mathbb{B}_{i,j} \big(\tau_i \otimes \tau_j\big) \right),
\end{equation}
the density matrix $\varrho_{\text{B}\bar{\text{B}}}$ thus takes a symmetric X-state form, which we label $\varrho^{\text{X}}_{\text{B}\bar{\text{B}}}$. Consequently, the matrix $\mathcal{S}_{\mu\nu}$ is transformed into
\begin{equation}
\mathcal{S}_{\mu\bar{\nu}} =
\begin{pmatrix}
1 & 0 & 0 & \mathbb{A}_{0,z} \\
0 & \mathbb{B}_{x,x} & 0 & 0 \\
0 & 0 & \mathbb{B}_{y,y} & 0 \\
\mathbb{A}_{z,0} & 0 & 0 & \mathbb{B}_{z,z}
\end{pmatrix}, \quad \text{where} \quad
\begin{aligned}
\mathbb{A}_{z,0}&=\mathbb{A}_{0,z} = \frac{\beta_{\psi}\sin(\varphi)\cos(\varphi)}{1+\alpha_{\psi}\cos^{2}(\varphi)}, \\
\mathbb{B}_{x,x} &= \frac{1 + \alpha_{\psi} + \sqrt{( \alpha_{\psi} +\cos2\varphi)^2 + \gamma_{\psi}^2 \sin^22\varphi}}{2(1+\alpha_{\psi}\cos^{2}\varphi)}, \\
\mathbb{B}_{y,y} &= \frac{1 + \alpha_{\psi} - \sqrt{( \alpha_{\psi} +\cos2\varphi)^2 + \gamma_{\psi}^2 \sin^22\varphi}}{2(1+\alpha_{\psi}\cos^{2}\varphi)}, \\
\mathbb{B}_{z,z} &= \frac{-\alpha_{\psi}\sin^{2}(\varphi)}{1 + \alpha_{\psi}\cos^{2}(\varphi)},
\end{aligned}
\end{equation}
where $\mathbb{A}_{z,0}$ represents the equal polarizations $\text{P}_{\text{B}} = \text{P}_{\bar{\text{B}}}$, and $\mathbb{B}_{z,z}$ corresponds to $\mathcal{S}_{yy}$. The parameters $\mathbb{B}_1$ and $\mathbb{B}_2$ result from the diagonalization of the $x\text{-}y$ sector of the correlation matrix $\mathcal{S}_{\mu\nu}$. It is noteworthy that both the coordinate reorientation $(\mathbf{\hat{y}} \leftrightarrow \mathbf{\hat{z}})$ and the diagonalization of the $x\text{-}y$ block of $\mathcal{S}_{\mu\nu}$ are equivalent to applying a local unitary transformation to the state.
\begin{equation}
\varrho^{\text{X}}_{\text{B}\bar{\text{B}}} = (U_{\text{B}} \otimes U_{\bar{\text{B}}})\varrho_{\text{B}\bar{\text{B}}}(U_{\text{B}} \otimes U_{\bar{\text{B}}})^{\dagger},
\label{eq:8}
\end{equation}
with $\text{U}_{\text{B}}$ and $\text{U}_{\bar{\text{B}}}$ representing independent unitary transformations on the $\text{B}$ and $\bar{\text{B}}$ Hilbert spaces. Expanding Eq. (\ref{eq:8}) in the $\sigma_z$ basis yields the following form for the spin density operator 
\begin{equation}
\varrho_{\text{B}\bar{\text{B}}}^{\text{X}} =
\begin{pmatrix}
\varrho_{1,1}& 0 & 0 & \varrho_{1,4} \\
0 & \varrho_{2,2} & \varrho_{2,2} & 0 \\
0 & \varrho_{2,2} & \varrho_{2,2}& 0 \\
\varrho_{1,4} & 0 & 0 & \varrho_{4,4}
\end{pmatrix}, \quad \text{where} \quad
\begin{aligned}
 \varrho_{1,1}&= \frac{1}{4}\bigg(1+2\mathbb{A}_{z,0}+\mathbb{B}_{z,z}\bigg), \quad
\varrho_{1,4} &= \varrho_{4,1} = \frac{1}{4}\bigg(\mathbb{B}_{x,x}-\mathbb{B}_{y,y}\bigg), \\
\varrho_{2,2} &= \varrho_{3,3} = \frac{1}{4}\bigg(1-\mathbb{B}_{z,z}\bigg), \quad
\varrho_{2,3} &= \varrho_{3,2} = \frac{1}{4}\bigg(\mathbb{B}_{x,x}+\mathbb{B}_{y,y}\bigg), \\
\varrho_{4,4} &= \frac{1}{4}\bigg(1-2\mathbb{A}_{z,0}+\mathbb{B}_{z,z}\bigg).
\end{aligned}
\label{eq:varrho}
\end{equation}
In the high-energy limit of the $e^{+}e^{-} \to \text{B}\bar{\text{B}}$ process (where $\alpha_{\psi}=1$ and $\beta_{\psi}=\gamma_{\psi}=0$), the single-baryon vector polarization vanishes, $\mathbf{P}_{\text{B}}=0$, and the density matrix reduces to \citep{H1,H2}
\begin{equation}
\varrho_{\text{B}\bar{\text{B}}}^{\text{X}} =\frac{1}{2\big(1+\cos^{2}\varphi}
\begin{pmatrix}
\cos^{2}\varphi & 0 & 0 & \cos^{2}\varphi \\
0 & 1 & 1 & 0 \\
0 & 1 & 1 & 0 \\
\cos^{2}\varphi & 0 & 0 & \cos^{2}\varphi
\end{pmatrix},
\end{equation}
In the low-energy (LE) limit close to threshold ($\alpha_{\psi}=\beta_{\psi}=0$, $\gamma_{\psi}=1$), the baryon polarization simplifies to the initial polarization of the electron beam
\begin{equation}
\varrho_{\text{B}\bar{\text{B}}}^{\text{X}} =\frac{1}{4}
\begin{pmatrix}
1 & 0 & 0 & 1 \\
0 & 1 & 1 & 0 \\
0 & 1 & 1 & 0 \\
1 & 0 & 0 & 1
\end{pmatrix},
\end{equation}
To compute the QFIM, we first evaluate the $16 \times 16$ matrix $\Lambda$ (Eq. \ref{landa}), for which the non-zero elements $\Lambda_{ij}$ are
\begin{equation}
\Lambda =
\begin{pmatrix}
\Lambda_{11} & 0_{4\times 4} & 0_{4\times 4} & \Lambda_{14} \\
0_{4\times 4} & \Lambda_{22} & \Lambda_{23} & 0_{4\times 4} \\
0_{4\times 4} & \Lambda_{32} & \Lambda_{33} & 0_{4\times 4} \\
\Lambda_{41} & 0_{4\times 4} & 0_{4\times 4} & \Lambda_{44}
\end{pmatrix},
\label{Lambda}
\end{equation}
where the $4 \times 4$ submatrices $\Lambda_{ij}$ (for $i, j = 1, 2, 3, 4$) are given by

\[
\Lambda_{11} =
\begin{pmatrix}
2\varrho_{11} & 0 & 0 & \varrho_{14} \\
0 & \varrho_{11}+\varrho_{22} & \varrho_{22} & 0 \\
0 & \varrho_{22} & \varrho_{11}+\varrho_{22} & 0 \\
\varrho_{14} & 0 & 0 & \varrho_{11}+\varrho_{44}
\end{pmatrix},
\qquad
\Lambda_{14}=\Lambda_{41}=
\begin{pmatrix}
\varrho_{14} & 0 & 0 & 0 \\
0 & \varrho_{14} & 0 & 0 \\
0 & 0 & \varrho_{14} & 0 \\
0 & 0 & 0 & \varrho_{14}
\end{pmatrix}.
\]

\[
\Lambda_{22}=\Lambda_{33} =
\begin{pmatrix}
\varrho_{11}+\varrho_{22} & 0 & 0 & \varrho_{14} \\
0 & 2\varrho_{22} & \varrho_{22} & 0 \\
0 & \varrho_{22} & 2\varrho_{22} & 0 \\
\varrho_{14} & 0 & 0 & \varrho_{22}+\varrho_{44}
\end{pmatrix},
\qquad
\Lambda_{23}=\Lambda_{32} =
\begin{pmatrix}
\varrho_{22} & 0 & 0 & 0 \\
0 & \varrho_{22} & 0 & 0 \\
0 & 0 & \varrho_{22} & 0 \\
0 & 0 & 0 & \varrho_{22}
\end{pmatrix}.
\]

\[
\Lambda_{44} =
\begin{pmatrix}
\varrho_{11}+\varrho_{44} & 0 & 0 & \varrho_{14} \\
0 & \varrho_{22}+\varrho_{44} & \varrho_{22} & 0 \\
0 & \varrho_{22} & \varrho_{22}+\varrho_{44} & 0 \\
\varrho_{14} & 0 & 0 & 2\varrho_{44}
\end{pmatrix}.
\]
After calculating, the pseudo-inverse of the matrix $\Lambda$ from Eq. \ref{Lambda} is found to be
\begin{equation}
\Lambda^{+} =
\begin{pmatrix}
\Lambda^{+}_{11} & 0_{4\times 4} & 0_{4\times 4} & \Lambda^{+}_{14} \\
0_{4\times 4} & \Lambda^{+}_{22} & 0_{4\times 4} & 0_{4\times 4} \\
0_{4\times 4} & 0_{4\times 4} & \Lambda^{+}_{33} & 0_{4\times 4} \\
\Lambda^{+}_{41} & 0_{4\times 4} & 0_{4\times 4} & \Lambda^{+}_{44}
\end{pmatrix},
\tag{22}
\end{equation}
where
\[
\Lambda^{+}_{11} =
\begin{pmatrix}
\bar{\Lambda}_{1,1} & 0 & 0 & \bar{\Lambda}_{1,4} \\
0 & \bar{\Lambda}_{2,2} & 0 & 0 \\
0 & 0 & \bar{\Lambda}_{3,3} & 0 \\
\bar{\Lambda}_{4,1} & 0 & 0 & \bar{\Lambda}_{4,4}
\end{pmatrix},
\qquad
\Lambda^{+}_{14} = \Lambda^{+}_{41} =
\begin{pmatrix}
\bar{\Lambda}_{1,13} & 0 & 0 & \bar{\Lambda}_{1,16} \\
0 & 0 & 0 & 0 \\
0 & 0 & 0 & 0 \\
\bar{\Lambda}_{4,13} & 0 & 0 & \bar{\Lambda}_{4,16}
\end{pmatrix},
\]

\[
\Lambda^{+}_{44} =
\begin{pmatrix}
\bar{\Lambda}_{13,13} & 0 & 0 & \bar{\Lambda}_{13,16} \\
0 & \bar{\Lambda}_{14,14} & 0 & 0 \\
0 & 0 & \bar{\Lambda}_{15,15} & 0 \\
\bar{\Lambda}_{16,13} & 0 & 0 & \bar{\Lambda}_{16,16}
\end{pmatrix},
\qquad
\Lambda^{+}_{22} = \Lambda^{+}_{33} =
\begin{pmatrix}
\bar{\Lambda}_{5,5} & 0 & 0 & \bar{\Lambda}_{5,8} \\
0 & \bar{\Lambda}_{6,6} & 0 & 0 \\
0 & 0 & \bar{\Lambda}_{7,7} & 0 \\
\bar{\Lambda}_{8,5} & 0 & 0 & \bar{\Lambda}_{8,8}
\end{pmatrix},
\]
the non-zero elements $\bar{\Lambda}_{i,j}$ of the $16 \times 16$ matrix $\bar{\Lambda}$ (where $i$ and $j$ are the row and column indices, respectively). The equations group elements that share the same value. All remaining, unlisted matrix elements are equal to zero, i.e., $\bar{\Lambda}_{i,j}=0$ (see Appendix \ref{AppA}). Using the definition~(\ref{vecA}), one writes
\begin{equation}
\mathrm{vec}\!\left[\partial_{\alpha_{\psi}}\varrho\right]
=
\bigl(
\partial_{\alpha_{\psi}}\varrho_{1,1},\;
0,\;
0,\;
\partial_{\alpha_{\psi}}\varrho_{1,4},\;
0,\;
\partial_{\alpha_{\psi}}\varrho_{2,2},\;
\partial_{\alpha_{\psi}}\varrho_{2,2},\;
0,\;
\partial_{\alpha_{\psi}}\varrho_{2,2},\;
\partial_{\alpha_{\psi}}\varrho_{2,2},\;
0,\;
\partial_{\alpha_{\psi}}\varrho_{1,4},\;
0,\;
0,\;
\partial_{\alpha_{\psi}}\varrho_{4,4}
\bigr)^{T}.
\label{vecAlpha}
\end{equation}
and
\begin{equation}
\mathrm{vec}\!\left[\partial_{\varphi}\varrho\right]
=
\bigl(
\partial_{\varphi}\varrho_{1,1},\;
0,\;
0,\;
\partial_{\varphi}\varrho_{1,4},\;
0,\;
\partial_{\varphi}\varrho_{2,2},\;
\partial_{\varphi}\varrho_{2,2},\;
0,\;
\partial_{\varphi}\varrho_{2,2},\;
\partial_{\varphi}\varrho_{2,2},\;
0,\;
\partial_{\varphi}\varrho_{1,4},\;
0,\;
0,\;
\partial_{\varphi}\varrho_{4,4}
\bigr)^{T}.
\label{vecPhi}
\end{equation}
Employing~(\ref{Flanda}), the QFIM writes as
\begin{equation}
F =
\begin{pmatrix}
F_{\alpha_{\psi}\alpha_{\psi}} & F_{\alpha_{\psi}\varphi} \\
F_{\varphi\alpha_{\psi}} & F_{\varphi\varphi}
\end{pmatrix}
=
\begin{pmatrix}
2\,\mathrm{vec}[\partial_{\alpha_{\psi}}\varrho]^{T}\Lambda^{+}\mathrm{vec}[\partial_{\alpha_{\psi}}\varrho]
&
2\,\mathrm{vec}[\partial_{\alpha_{\psi}}\varrho]^{T}\Lambda^{+}\mathrm{vec}[\partial_{\varphi}\varrho]
\\[2mm]
2\,\mathrm{vec}[\partial_{\varphi}\varrho]^{T}\Lambda^{+}\mathrm{vec}[\partial_{\alpha_{\psi}}\varrho]
&
2\,\mathrm{vec}[\partial_{\varphi}\varrho]^{T}\Lambda^{+}\mathrm{vec}[\partial_{\varphi}\varrho]
\end{pmatrix}.
\label{QFIM}
\end{equation}
The pseudo-inverse of the QFIM given by
\begin{equation}
F^{-1}
=
\frac{1}{\det(F)}
\begin{pmatrix}
F_{\varphi\varphi} & -F_{\alpha\varphi} \\
-F_{\alpha_{\psi}\varphi} & F_{\alpha_{\psi}\alpha_{\psi}}
\end{pmatrix}.
\label{Fminus1}
\end{equation}
Therefore, from the inequality~\eqref{Fminus1},
\begin{align}
\mathrm{Var}(\alpha_{\psi}) &\ge \frac{F_{\varphi\varphi}}{\det(F)}, \\[1mm]
\mathrm{Var}(\varphi) &\ge \frac{F_{\alpha_{\psi}\alpha_{\psi}}}{\det(F)},
\end{align}
and
\begin{equation}
\Bigl(\mathrm{Var}(\alpha_{\psi}) - \frac{F_{\varphi\varphi}}{\det F}\Bigr)
\Bigl(\mathrm{Var}(\varphi) - \frac{F_{\alpha_{\psi}\alpha_{\psi}}}{\det F}\Bigr)
\;\ge\;
\Bigl(\mathrm{Cov}(\alpha_{\psi},\varphi) + \frac{F_{\alpha_{\psi}\varphi}}{\det F}\Bigr)^{2}.
\label{CRB_joint}
\end{equation}

Using~(\ref{vecL}), the SLD with respect to $\alpha_{\psi}$ is
\begin{equation}
L_{\alpha_{\psi}} = 2
\begin{pmatrix}
L^{\alpha_{\psi}}_{1,1} & 0 & 0 & L^{\alpha_{\psi}}_{1,4} \\
0 & L^{\alpha_{\psi}}_{2,2} & L^{\alpha_{\psi}}_{2,2} & 0 \\
0 & L^{\alpha_{\psi}}_{2,2} & L^{\alpha_{\psi}}_{2,2} & 0 \\
L^{\alpha_{\psi}}_{1,4} & 0 & 0 & L^{\alpha_{\psi}}_{4,4}
\end{pmatrix}.
\label{SLDalpha}
\end{equation}

\begin{align*}
L^{\alpha_{\psi}}_{1,1} &=
\frac{
 \varrho_{44}^{2}\,\partial_{\alpha_{\psi}}\varrho_{11}
 - \varrho_{14}^{2}\,\partial_{\alpha_{\psi}}\varrho_{11}
 + \varrho_{14}^{2}\,\partial_{\alpha_{\psi}}\varrho_{44}
 + \varrho_{11}\varrho_{44}\,\partial_{\alpha_{\psi}}\varrho_{11}
 - 2\varrho_{14}\varrho_{44}\,\partial_{\alpha_{\psi}}\varrho_{14}}{(\varrho_{11}+\varrho_{44})(\varrho_{11}\varrho_{44}-\varrho_{14}^{2})},\\[2mm]
L^{\alpha_{\psi}}_{1,4} &=\frac{\varrho_{44}(2\varrho_{11}\partial_{\alpha_{\psi}}\varrho_{14}-\varrho_{14}\partial_{\alpha_{\psi}}\varrho_{11})-\varrho_{11}\varrho_{14}\,\partial_{\alpha_{\psi}}\varrho_{44}}{(\varrho_{11}+\varrho_{44})(\varrho_{11}\varrho_{44}-\varrho_{14}^{2})},\\[2mm]
L^{\alpha_{\psi}}_{2,2} &= \frac{\partial_{\alpha_{\psi}}\varrho_{22}}{2\varrho_{22}},
\\[2mm]
L^{\alpha_{\psi}}_{4,4} &=\frac{\varrho_{14}^{2}\partial_{\alpha_{\psi}}\varrho_{11}
 + \varrho_{11}^{2}\partial_{\alpha_{\psi}}\varrho_{44}
 - \varrho_{14}^{2}\partial_{\alpha_{\psi}}\varrho_{44}
 - 2\varrho_{11}\varrho_{14}\partial_{\alpha_{\psi}}\varrho_{14}
 + \varrho_{11}\varrho_{44}\partial_{\alpha_{\psi}}\varrho_{44}
}{(\varrho_{11}+\varrho_{44})(\varrho_{11}\varrho_{44}-\varrho_{14}^{2})}.
\end{align*}
and
\begin{equation}
L_{\varphi} = 2
\begin{pmatrix}
L^{\varphi}_{1,1} & 0 & 0 & L^{\varphi}_{1,4} \\
0 & L^{\varphi}_{2,2} & L^{\varphi}_{2,2} & 0 \\
0 & L^{\varphi}_{2,2} & L^{\varphi}_{2,2} & 0 \\
L^{\varphi}_{1,4} & 0 & 0 & L^{\varphi}_{4,4}
\end{pmatrix}
\end{equation}
where
\begin{align*}
L^{\varphi}_{1,1} &=
\frac{
 \varrho_{44}^{2}\,\partial_{\varphi}\varrho_{11}
 - \varrho_{14}^{2}\,\partial_{\varphi}\varrho_{11}
 + \varrho_{14}^{2}\,\partial_{\varphi}\varrho_{44}
 + \varrho_{11}\varrho_{44}\,\partial_{\varphi}\varrho_{11}
 - 2\varrho_{14}\varrho_{44}\,\partial_{\varphi}\varrho_{14}
}{(\varrho_{11}+\varrho_{44})(\varrho_{11}\varrho_{44}-\varrho_{14}^{2})},\\[3mm]
L^{\varphi}_{1,4} &=\frac{\varrho_{44}\bigl(2\varrho_{11}\partial_{\varphi}\varrho_{14}-\varrho_{14}\partial_{\varphi}\varrho_{11}\bigr)- \varrho_{11}\varrho_{14}\,\partial_{\varphi}\varrho_{44}}{(\varrho_{11}+\varrho_{44})(\varrho_{11}\varrho_{44}-\varrho_{14}^{2})},\\[3mm]
L^{\varphi}_{2,2} &= \frac{\partial_{\varphi}\varrho_{22}}{2\varrho_{22}},\\[3mm]
L^{\varphi}_{4,4} &=\frac{\varrho_{14}^{2}\,\partial_{\varphi}\varrho_{11}+ \varrho_{11}^{2}\,\partial_{\varphi}\varrho_{44}
 - \varrho_{14}^{2}\,\partial_{\varphi}\varrho_{44}
 - 2\varrho_{11}\varrho_{14}\,\partial_{\varphi}\varrho_{14}
 + \varrho_{11}\varrho_{44}\,\partial_{\varphi}\varrho_{44}
}{(\varrho_{11}+\varrho_{44})(\varrho_{11}\varrho_{44}-\varrho_{14}^{2})}.
\end{align*}
When the first two inequalities are saturated, the resulting precision for the parameters $\alpha_{\psi}$ and $\varphi$ is maximized. The corresponding minimal values of the variances are therefore
\begin{equation}
\begin{aligned}
\mathrm{Var}(\varphi)_{\min} &= \frac{F_{\alpha_{\psi}\alpha_{\psi}}}{\det(F)}, 
\end{aligned}
\label{Varvarphi}
\end{equation}

\begin{equation}
\begin{aligned}
\mathrm{Var}(\alpha_{\psi})_{\min} &= \frac{F_{\varphi\varphi}}{\det(F)}.
\end{aligned}
\label{Varalpha}
\end{equation}

By assuming statistical independence, we ensure the parameters remain uncorrelated; thus, the determination of one parameter does not affect the precision of the others. This condition is satisfied if and only if the QFIM is diagonal ($F_{ij} = 0$ for $i \neq j$), implying
\begin{equation}
{\rm Var}\left( \varphi  \right)^{\rm Ind}_{\rm min} \ge F_{\varphi \varphi }^{ - 1},
\label{varind}
\end{equation}

\begin{equation}
{\rm Var}\left( \alpha_{\psi}  \right)^{\rm Ind}_{\rm min} \ge F_{\alpha_{\psi}\alpha_{\psi}}^{ - 1} \label{varind1}.
\end{equation}
\section{Results and discussions}
Utilizing data provided by the BESIII collaboration, the following table details the fundamental parameters relevant to the $J/\psi \to \text{B}\bar{\text{B}}$ production and decay chain.
\begin{table}[H]
\centering
\caption{Parameters characterizing the production of ground-state octet hyperon pairs ($\text{B} \bar{\text{B}}$) via the $J/\psi$ resonance in $e^{+}e^{-}$ annihilation.}
 \label{t1}
\begin{tabular}{c c c c c}
\hline
\hline

$\text{B}\bar{\text{B}}$ & $\Lambda\bar{\Lambda}$ & $\Sigma^{+}\bar{\Sigma}^{-}$ &  $\Xi^{-}\bar{\Xi}^{+}$ &  $\Xi^{0}\bar{\Xi}^{0}$\\

\hline

$\alpha_{\psi}$ & 0.475(4) & -0.508(7)  & 0.586(16)& 0.514(16) \\
\hline
$\Delta\Phi/rad$ & 0.752(8) & -0.270(15) & 1.213(49) & 1.168(26)\\
\hline
References & \cite{ref61,T2} & \cite{T3,T4} & \cite{ref56,T6} & \cite{ref63,reff66} \\
\hline
\hline
\end{tabular}
\label{tab1}
\end{table}

\subsection{QFI in the abscence of noisy channels}\label{sec:4}
The subsequent investigation is dedicated to the precise estimation of the scattering angle $\varphi$ and the decay parameter $\alpha_{\psi}$ associated with the vector charmonium $\psi(c\bar{c})$.

\begin{figure}[!h]
\includegraphics[scale=0.4]{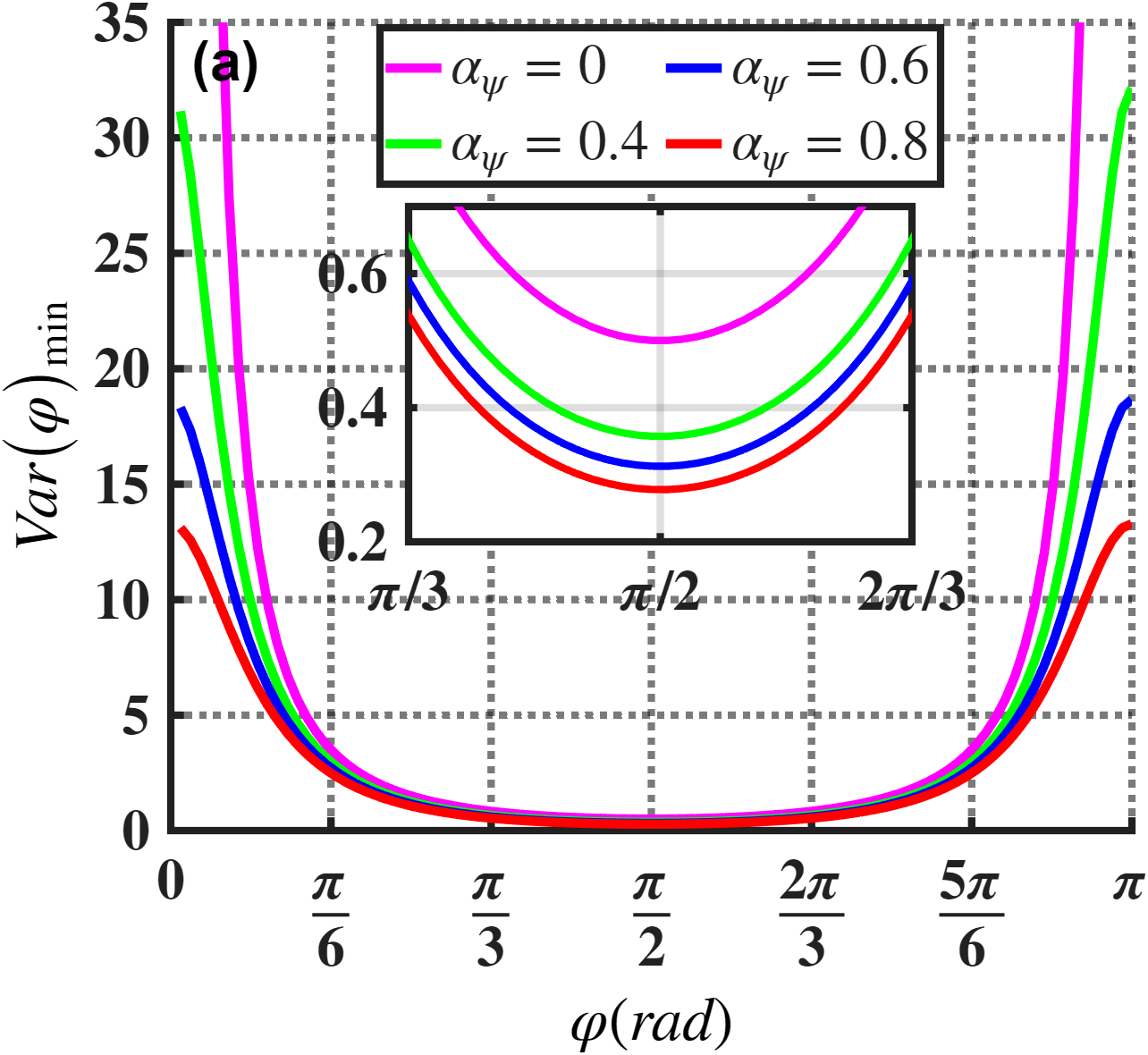}
\includegraphics[scale=0.4]{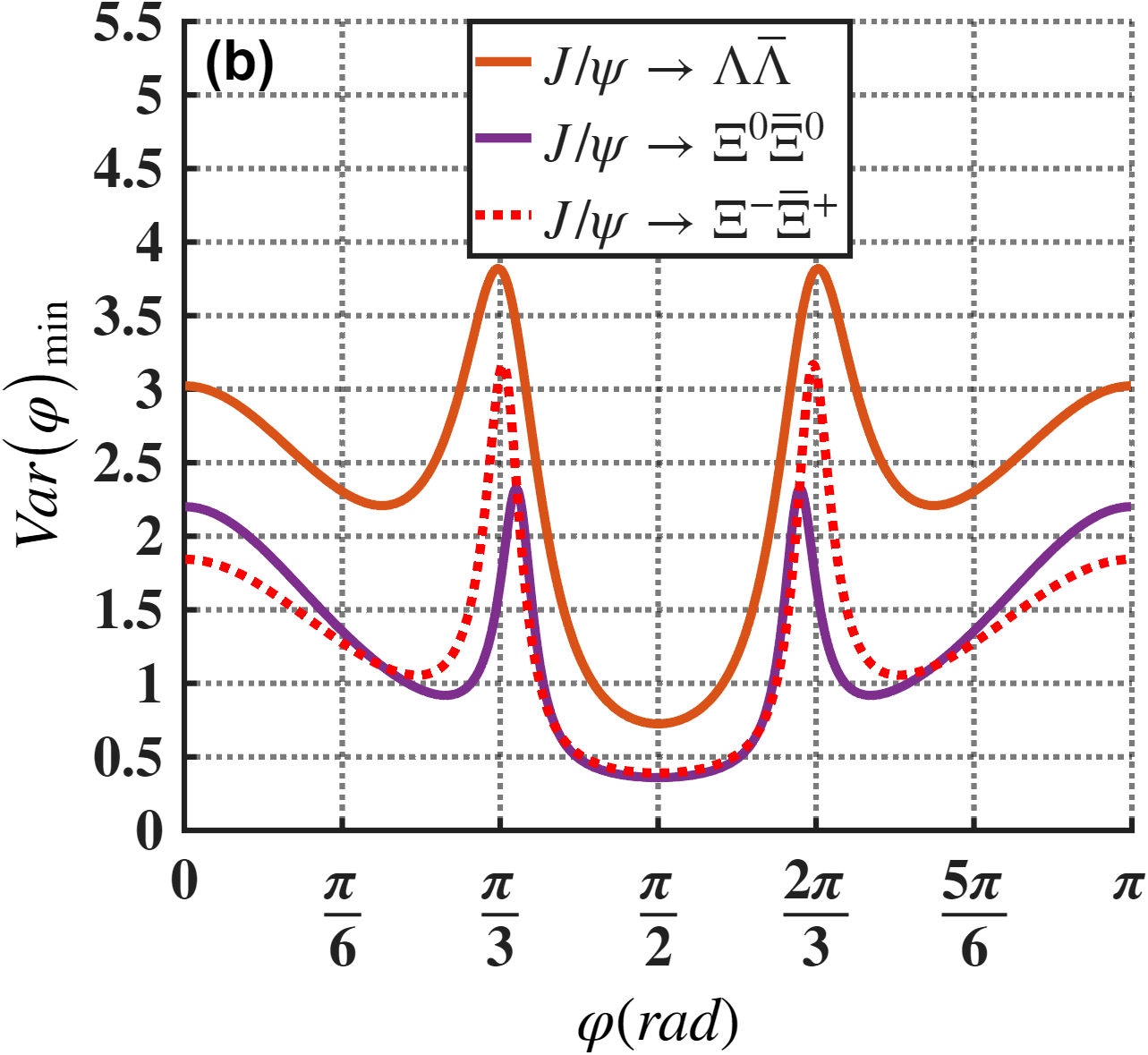}
\caption{(a) Plot of the minimum variances of the simultaneous estimates $\mathrm{Var}(\varphi)_{\min}$ in the $e^+e^- \to J/\psi \to \text{B} \bar{\text{B}}$ process as a function of the scattering angle $\varphi$ in theoretical case: $\alpha_{\psi} \in [0,1]$, $\beta_{\psi}=0.4$ and $\gamma_{\psi}=0$. (b) Plot of the minimum variances of the simultaneous estimates $\mathrm{Var}(\varphi)_{\min}$ for the $e^+e^- \to J/\psi \to \text{B} \bar{\text{B}}$ process, with $ \text{B} \bar{\text{B}}\equiv (\Lambda\bar{\Lambda}, \Xi^0\bar{\Xi}^0, \Xi^-\bar{\Xi}^+)$ channels, incorporating the experimental values provided in Table~\ref{t1}.}%
\label{fig:varphi1}
\end{figure}

\begin{figure}[!h]
\includegraphics[scale=0.4]{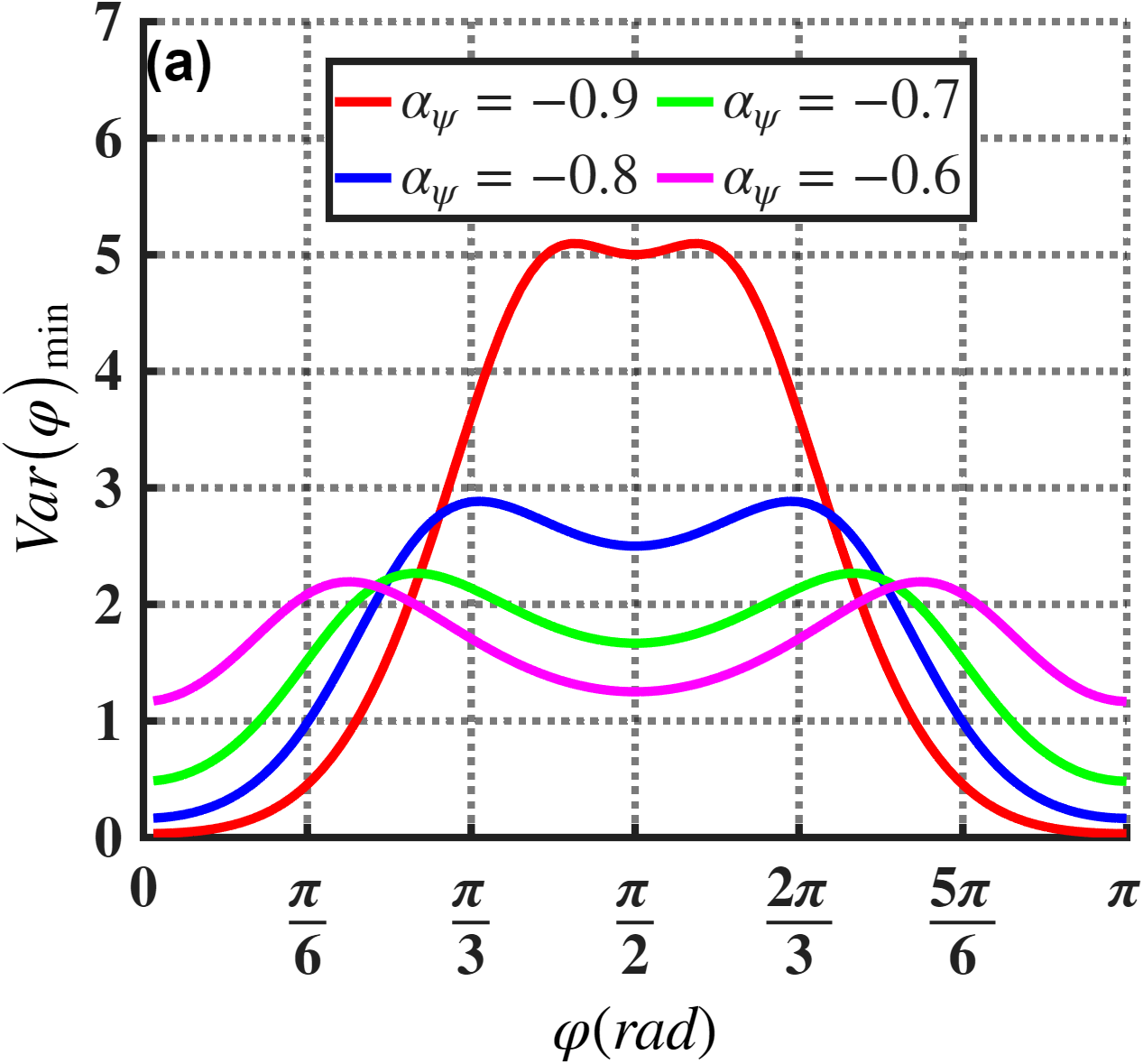}
\includegraphics[scale=0.4]{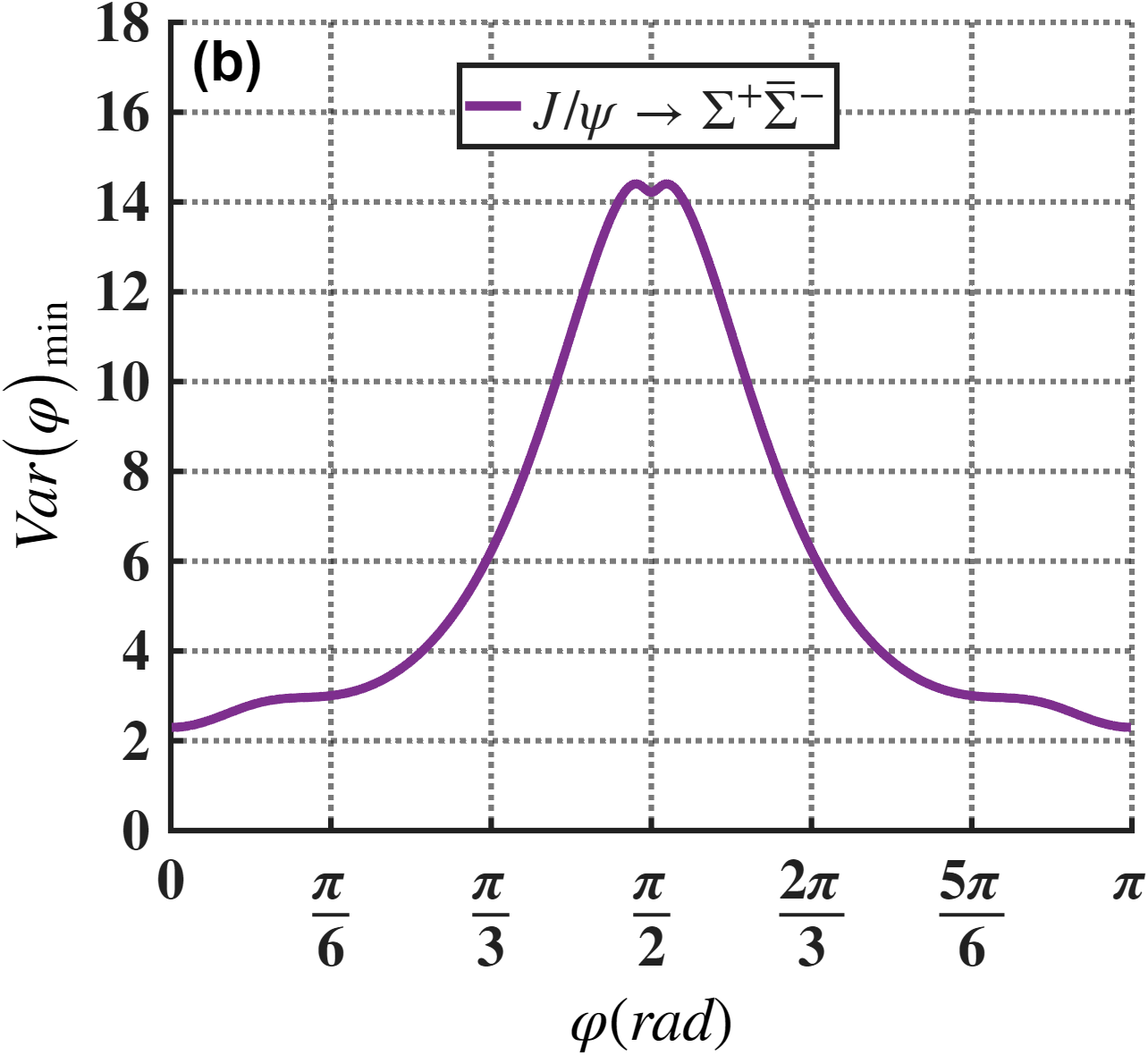}
\caption{(a) Plot of the minimum variances of the simultaneous estimates $\mathrm{Var}(\varphi)_{\min}$ in the $e^+e^- \to J/\psi \to \text{B} \bar{\text{B}}$ process as a function of the scattering angle $\varphi$ in theoretical case: $\alpha_{\psi} \in [-1,0]$, $\beta_{\psi}=0.4$ and $\gamma_{\psi}=0$. (b) Plot of the minimum variances of the simultaneous estimates $\mathrm{Var}(\varphi)_{\min}$ for the $e^+e^- \to J/\psi \to \text{B} \bar{\text{B}}$ process, with $ \text{B} \bar{\text{B}}\equiv (\Lambda\bar{\Lambda}, \Xi^0\bar{\Xi}^0, \Xi^-\bar{\Xi}^+)$ channels, incorporating the experimental values provided in Table~\ref{t1}.}%
\label{fig:varphi2}
\end{figure}

The foremost aim of the investigation presented in Figs.~\ref{fig:varphi1} and \ref{fig:varphi2} is to examine the minimum variance $\mathrm{Var}(\varphi)_{\min}$ as a function of the parameter $\varphi$. Figures \ref{fig:varphi1}(a) and \ref{fig:varphi2}(a) correspond to the parameter ranges $\alpha_{\psi} \in [0, 1]$ and $\alpha_{\psi} \in [-1, 0]$, respectively. In contrast, Figs. \ref{fig:varphi1}(b) and \ref{fig:varphi2}(b) illustrate the results using the specific values of $\alpha_{\psi}$ for the different hyperon decay channels.

We note that for $\alpha_{\psi} \in [0,1]$, the minimum variance $\mathrm{Var}(\varphi)_{\min}$ is smallest near the optimal angle $\varphi \to \pi/2$. This reduction in $\mathrm{Var}(\varphi)_{\min}$ indicates maximum estimation precision, where the QFI $F$ reaches its upper bound. Conversely, $\mathrm{Var}(\varphi)_{\min}$ increases monotonically as $\varphi$ deviates from $\pi/2$, reaching its maximum value as $\varphi \to \{0, \pi\}$, which corresponds to minimum estimation precision.

The results shown in Fig.~\ref{fig:varphi1}(a) are in good agreement with the variance expressed in Eq.~(\ref{Varvarphi}) when setting $\beta_{\psi}=\gamma_{\psi}=0$. By further fixing the decay parameter to $\alpha_{\psi}=1$, one obtains
\begin{equation}
\mathrm{Var}(\varphi)_{\min} = \frac{(1+\cos^2\upsilon)^2}{4\,\sin^2\upsilon},
\end{equation}
as shown by this expression, the minimum variance diverges at the boundaries $\varphi \in \{0, \pi\}$ and vanishes at $\varphi = \varphi^{\rm opt} = \pi/2$, indicating that
\begin{equation*} 
\text{Var}(\varphi)_{\min} =
\begin{cases}
0.25, & \text{if } \varphi = \pi/2, \\[6pt]
+\infty, & \text{if } \varphi = 0, \pi.
\end{cases}
\end{equation*}

As \(\alpha_{\psi}\to -1\), the minimal variance \(\mathrm{Var}(\varphi)_{\min}\) vanishes at the extreme of \(\varphi\) while reaching high values near \(\varphi \to \pi/2\), as depicted in Fig.~\ref{fig:varphi2}(a). The optimal scattering angles are thus \(\varphi=0,\pi\), which means the maximum precision, corresponding to the minimal variance \(\mathrm{Var}(\varphi)_{\min}\), is achieved for \(\alpha_{\psi} \to -1\). This optimum is illustrated for the process \(e^+e^- \to J/\psi \to \Sigma^+\bar{\Sigma}^-\) in Fig.~\ref{fig:varphi2}(b), highlighting the characteristic dependence of $\mathrm{Var}(\varphi)_{\min}$ on $\varphi$ and $\alpha_{\psi}$. Moreover, at the optimal angle $\varphi = \pi/2$, the minimal variance $\mathrm{Var}(\varphi)_{\min}$ is independent of $\gamma_{\psi}$, and Eq.~(\ref{Varvarphi}) reduces to
\begin{equation} \label{F3a}
\mathrm{Var}(\varphi)_{\min}=\frac{1-\alpha_{\psi}}{2\beta_{\psi}^{2}},
\end{equation}
as shown in Eq.~(\ref{F3a}), the minimum variance $\mathrm{Var}(\varphi)_{\min}$ peaks at $\alpha_{\psi}=-1$ and falls to zero at $\alpha_{\psi}=1$.
\begin{equation*} 
\text{Var}(\varphi)_{\min} =
\begin{cases}
0, & \text{if } \alpha_{\psi} = 1, \\
1/\beta_{\psi}^{2}, & \text{if }  \alpha_{\psi}= -1.
\end{cases}
\end{equation*}
In the high-energy limit of the $e^{+}e^{-} \to \mathrm{B}\bar{\mathrm{B}}$ process (where $\alpha_{\psi}=1$ and $\beta_{\psi}=\gamma_{\psi}=0$), the single-baryon vector polarization vanishes, i.e., $\mathbf{P}_{\mathrm{B}}=0$ \citep{H1,H2}. Consequently, the $\mathrm{B}\bar{\mathrm{B}}$ state is dominated by strong spin-spin correlations, yielding a highly entangled baryon-antibaryon pair and maximal quantum Fisher information \citep{JLAM}. This enables optimal metrological sensitivity in the high-energy regime.

\begin{figure}[!h]
\includegraphics[scale=0.4]{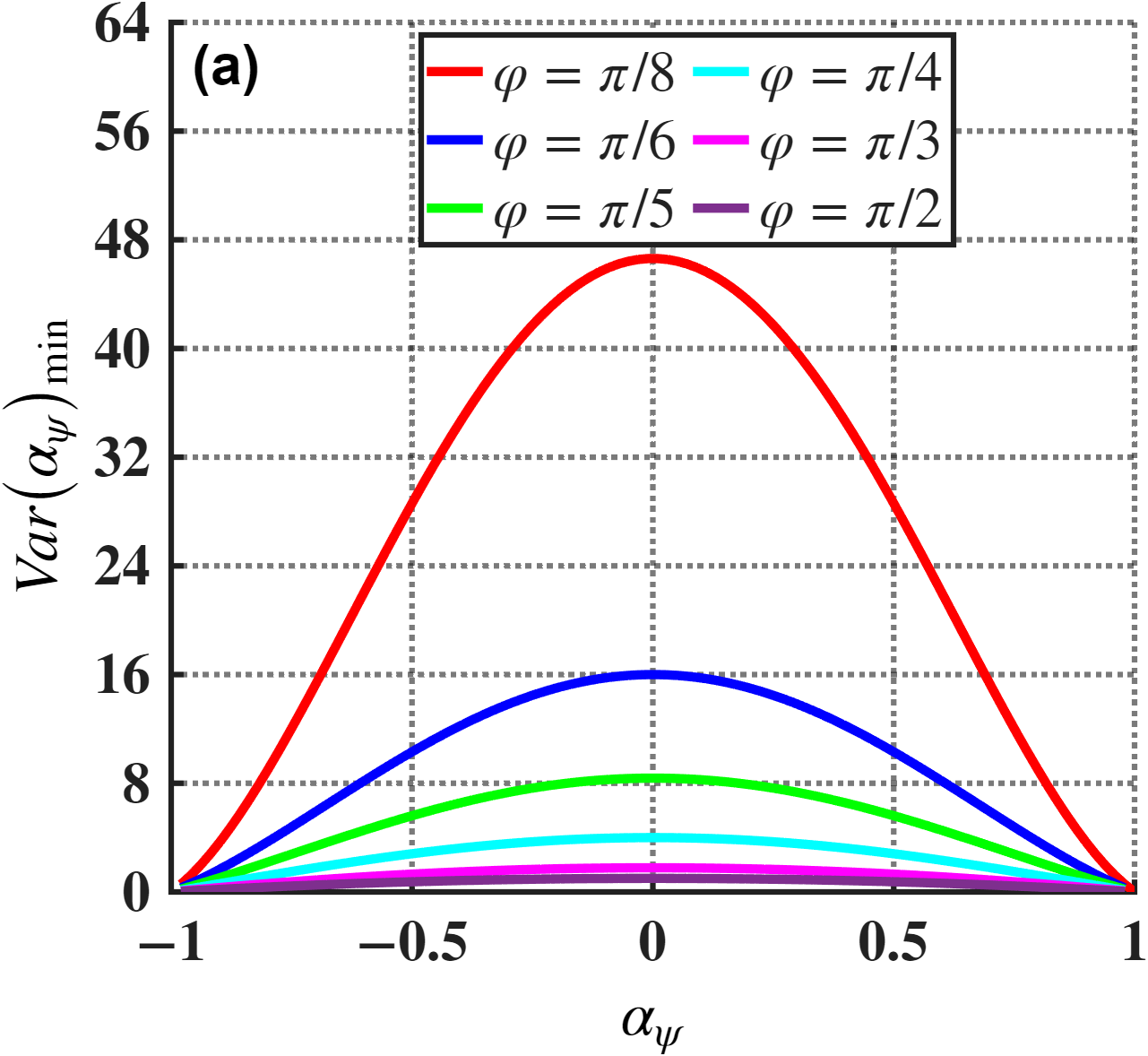}
\includegraphics[scale=0.4]{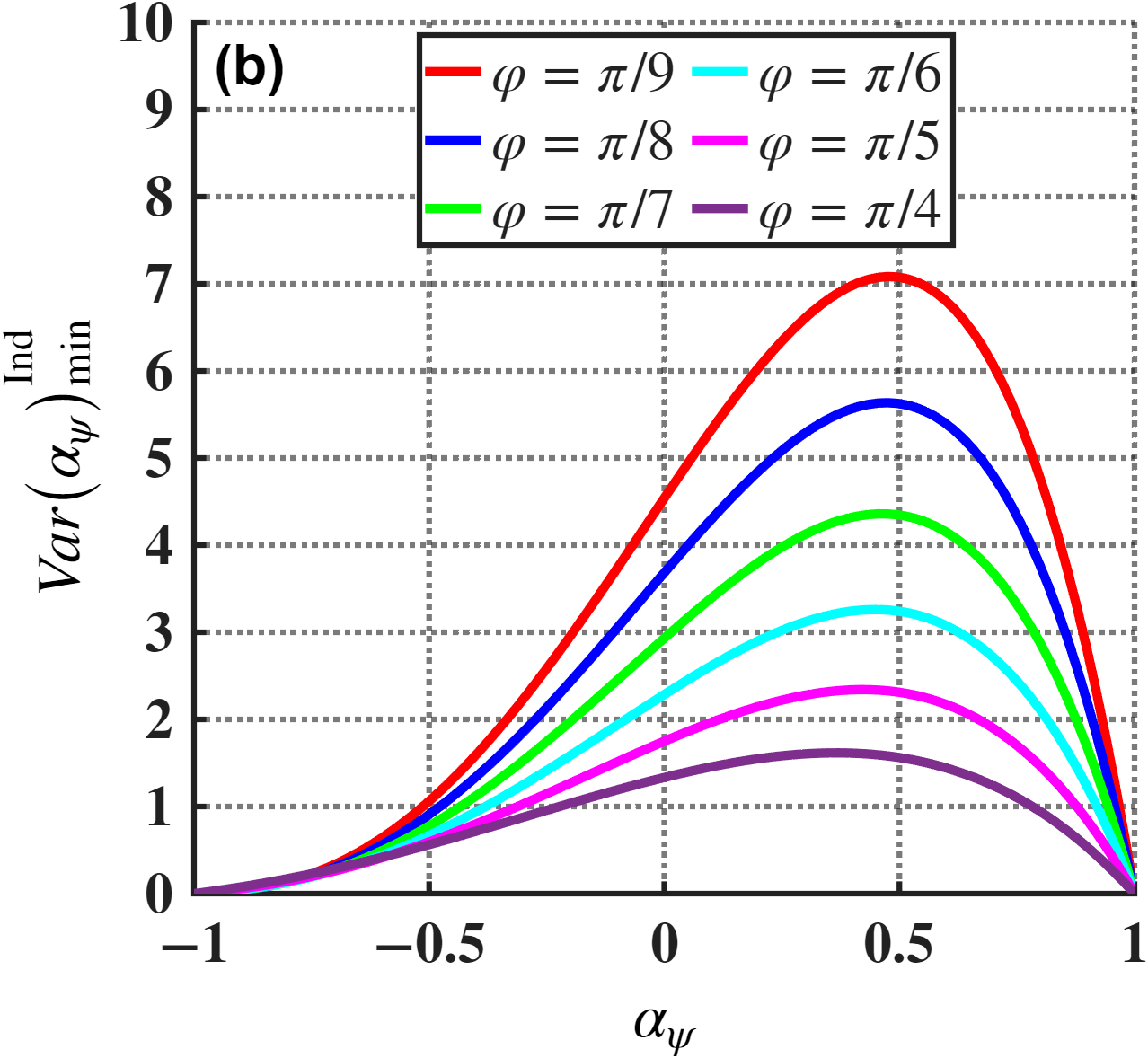}
\caption{(a) Plot of the minimal variance of the simultaneous estimation $\mathrm{Var}(\alpha_{\psi})_{\min}$ of the decay parameter $\alpha_{\psi}$ as a function of the $\alpha_{\psi}$ for various value of the angle $\varphi$. (b) Plot of the minimal variance for the individual estimation of the decay parameter $\alpha_{\psi}$ as a function of the $\alpha_{\psi}$ for various value of the $\varphi$, in the high-energy limit ($\beta_{\psi}=\gamma_{\psi}=0$).}
\label{fig:alpha1}
\end{figure}
In Fig.~\ref{fig:alpha1}(a), we present the influence of the $\alpha_{\psi}$ and the angle $\varphi$ on the evolution of the $\mathrm{Var}(\alpha_{\psi})_{\min}$. We remark that this variance vanishes for $\alpha_{\psi}\to -1$. It can be seen that high precision in the estimation of $\alpha_{\psi}$ is realized as $\varphi \to \pi/2$, as shown in Fig.~\ref{fig:alpha1}(a); these results are consistent with the previous figures. Moreover, $\mathrm{Var}(\alpha_{\psi})_{\min}$ peaks at $\alpha_{\psi}=0$ for a fixed $\varphi$ and vanishes as $\alpha_{\psi} \to \pm 1$, as illustrated in Fig.~\ref{fig:alpha1}(a). This reflects an increase in the Local Quantum Fisher Information (LQFI) and stronger entanglement.

The findings in Fig.~\ref{fig:alpha1}(a) are a direct consequence of the variance in Eq.~(\ref{Varalpha}). For the specific case $\varphi=\pi/2$, $\mathrm{Var}(\alpha_{\psi})_{\min}$ becomes independent of both $\beta_{\psi}$ and $\gamma_{\psi}$, simplifying to
\begin{equation} \label{Eqalpha1}
\mathrm{Var}(\alpha_{\psi})_{\min} = 1 - \alpha_{\psi}^2,
\end{equation}
with $\mathrm{Var}(\alpha_{\psi})_{\min}$ reaching its maximum at $\alpha_{\psi}=0$ and vanishing at the boundaries $\alpha_{\psi}=\pm 1$, Eq. (\ref{Eqalpha1}) reduces to
\begin{equation*}
\mathrm{Var}(\alpha_{\psi})_{\min} =
\begin{cases}
0, & \text{if } \alpha_{\psi} = \pm 1, \\
1, & \text{if } \alpha_{\psi} = 0.
\end{cases}
\end{equation*}
When $\alpha_{\psi} = \beta_{\psi} = \gamma_{\psi} = 0$, the minimal variance varies with $\upsilon$ according to
\begin{equation} \label{Eqalpha2}
\mathrm{Var}(\alpha_{\psi})_{\min} = \frac{1}{(1 - \cos^2 \upsilon)^2},
\end{equation}
reaching a maximum for $\varphi = 0, \pi$ and a minimum for $\varphi = \pi/2$, Eq. (\ref{Eqalpha2}) became
\begin{equation*}
\mathrm{Var}(\alpha_{\psi})_{\min} =
\begin{cases}
1, & \text{if } \varphi = \pi/2, \\
+\infty, & \text{if } \varphi = 0, \pi.
\end{cases}
\end{equation*}
The following analysis focuses on the independent estimation of parameters. The results shown in Fig.~\ref{fig:alpha1}(b) for $\mathrm{Var}(\alpha_{\psi})_{\min}$ are fully consistent with those obtained in the simultaneous-estimation scheme [Fig.~\ref{fig:alpha1}(a)]. Moreover, as in the simultaneous case, the variance decreases as $\varphi\to\pi/2$, indicating an enhanced precision in estimating $\alpha_{\psi}$, where the associated quantum Fisher information is maximal. Conversely, at the poles $\varphi=0,\pi$ and for $\alpha_{\psi}=0$, the variance peaks, signaling a decrease in precision and a corresponding minimum in the QFI.
\begin{figure}[!h]
\includegraphics[scale=0.4]{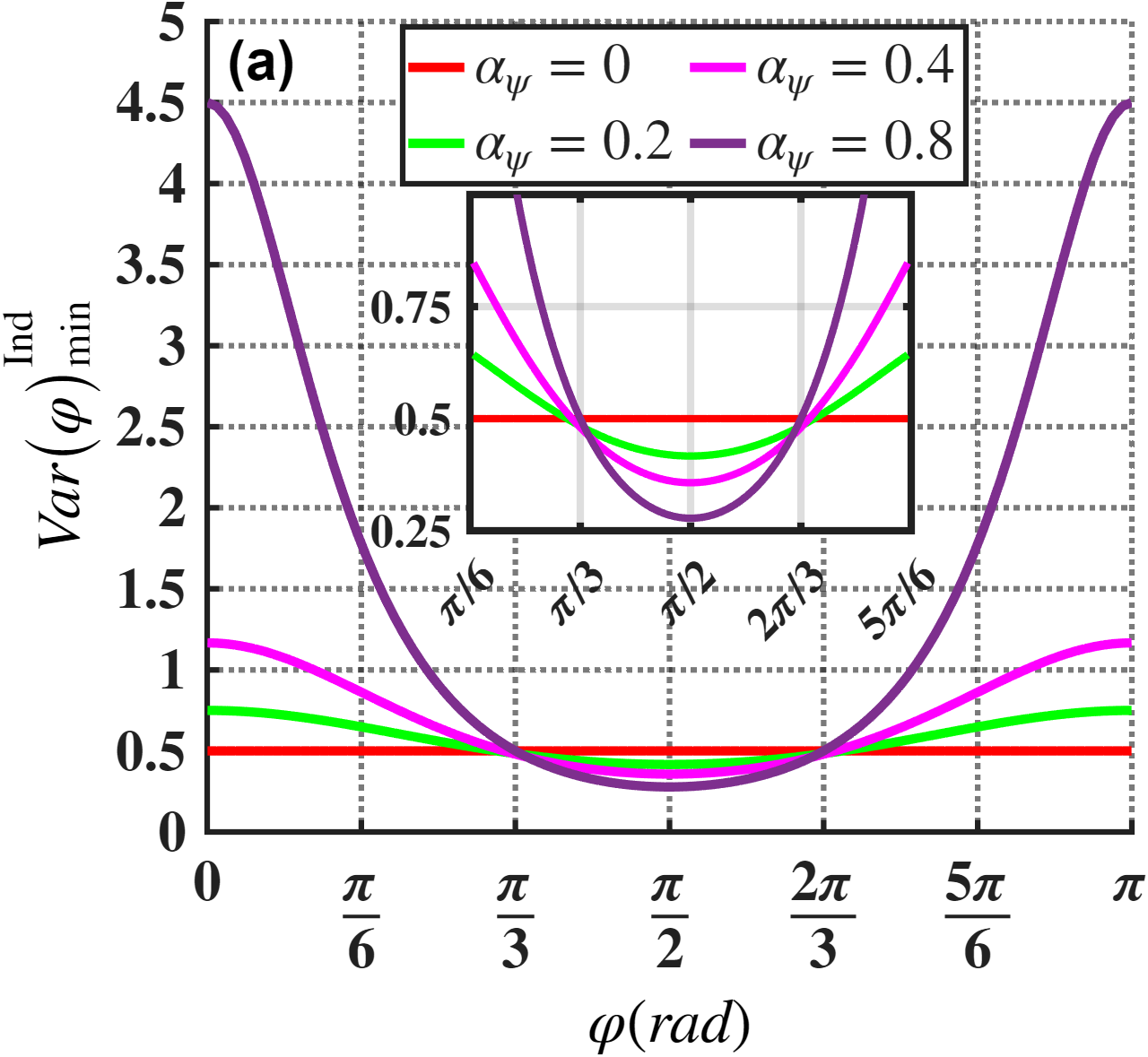}
\includegraphics[scale=0.4]{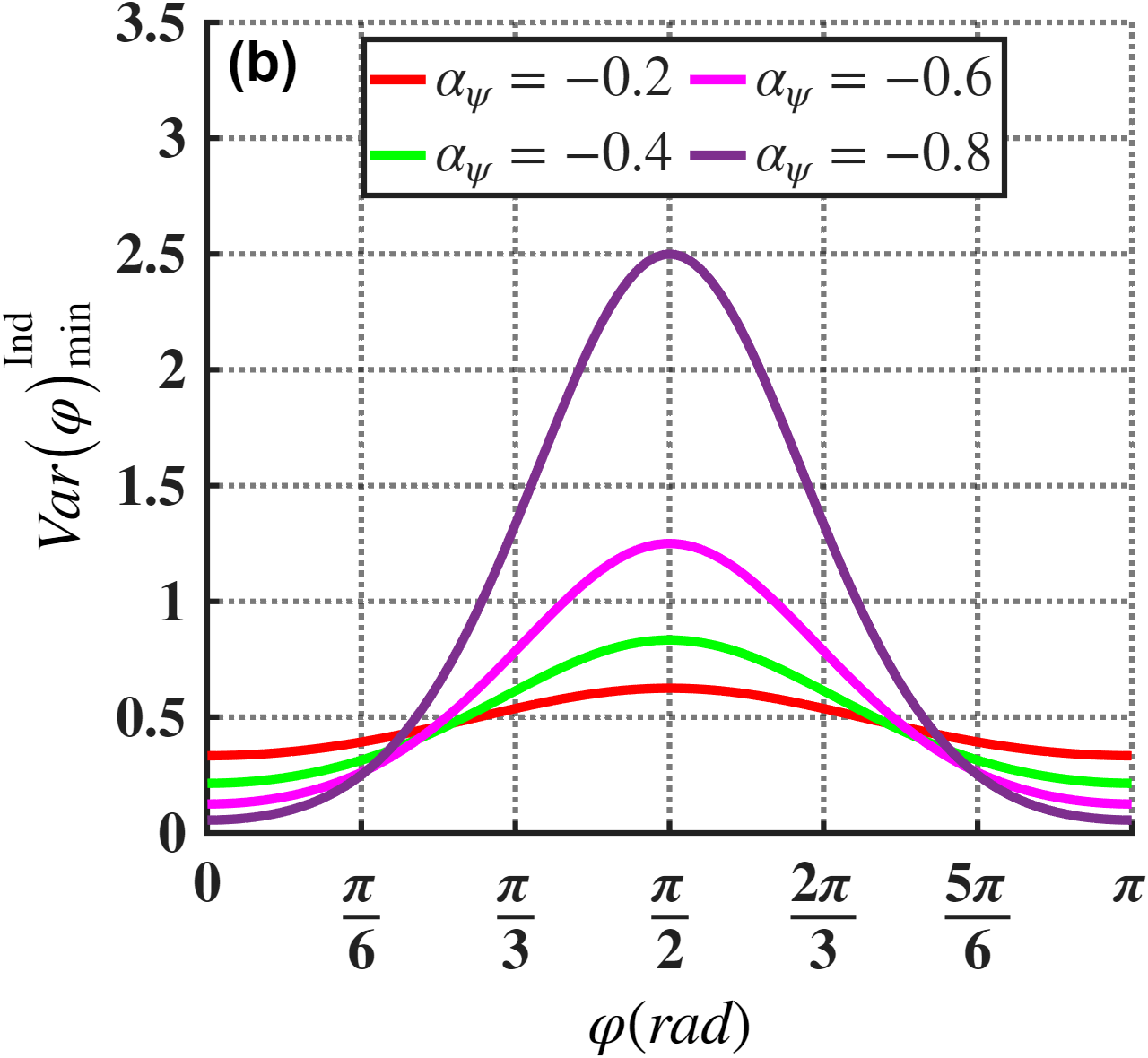}
\caption{Plot of the minimum variance of Individual estimates ${\rm Var}(\varphi)^{\rm Ind}_{\rm min}$ of the scattering angle $\varphi$ with $\beta_{\psi}=0.4$ and $\gamma_{\psi}=0$, for (a) $\alpha_{\psi}\in[0,1]$, and (b) $\alpha_{\psi}\in[-1,0]$.}
\label{fig:Varvarphi2}
\end{figure}

In Fig.~\ref{fig:Varvarphi2}(a), we present the minimal variance for the individual estimation ${\rm Var}(\varphi)^{\rm Ind}_{\rm min}$ of the scattering angle $\varphi$ for $\alpha_{\psi}\in[0,1]$, with $\beta_{\psi}=\gamma_{\psi}=0$. The figure illustrates that ${\rm Var}(\varphi)^{\rm Ind}_{\rm min}$ is maximized at $\varphi=\{ 0, \pi\}$ as $\alpha_{\psi}\to 1$. In contrast, the variance reaches its minimum near $\varphi = \pi/2$, where high estimation precision is achieved. Furthermore, for $\alpha_{\psi}=\gamma_{\psi}=0$, the variance becomes independent of $\beta_{\psi}$ and remains constant at $1/2$. By fixing the scattering angle to $\varphi=0, \pi/2,$ or $\pi$, Eq.~(\ref{varind}) becomes independent of $\gamma_{\psi}$ and reduces to
\begin{equation}
{\rm Var}(\varphi)^{\rm Ind}_{\rm min}=
\begin{cases}
\dfrac{1-\alpha_{\psi}}{2\beta_{\psi}^{2}}, & \text{if }  \varphi =\pi/2, \\[6pt]
\dfrac{(1+\alpha_{\psi})^{2}}{2\beta_{\psi}^{2}}, & \text{if } \varphi = 0,\pi.
\end{cases}
\label{Ind}
\end{equation}
Furthermore, as depicted in Fig.~\ref{fig:Varvarphi2}(a), ${\rm Var}(\varphi)^{\rm Ind}_{\rm min}$ increases monotonically with $\alpha_{\psi}$ at $\varphi=0$ and $\pi$, while it decreases monotonically with $\alpha_{\psi}$ at $\varphi=\pi/2$.

Figure~\ref{fig:Varvarphi2}(b) displays the minimal variance ${\rm Var}(\varphi)^{\rm Ind}_{\rm min}$ for the individual estimation of the scattering angle $\varphi$ with $\alpha_{\psi}\in[-1,0]$ and $\beta_{\psi}=\gamma_{\psi}=0$. In this case, the behavior observed in Fig.~\ref{fig:Varvarphi2}(a) is reversed. We find that ${\rm Var}(\varphi)^{\rm Ind}_{\rm min}$ reaches its maximum at $\varphi=\pi/2$ as $\alpha_{\psi}\to -1$, whereas it is minimized at $\varphi=0$ and $\pi$ (where estimation precision is maximal). Furthermore, based on Eq.~(\ref{Ind}), the minimal variance at $\varphi=\pi/2$ reduces to
\begin{equation*}
{\rm Var}(\varphi)^{\rm Ind}_{\rm min}=\frac{1-\alpha_{\psi}}{2\beta_{\psi}^{2}}=
\begin{cases}
0, & \text{if } \alpha_{\psi}=1, \\[6pt]
1/\beta_{\psi}^{2}, & \text{if }  \alpha_{\psi}=-1,
\end{cases}
\end{equation*}
while for $\varphi=0$ or $\pi$ one obtains
\begin{equation*}
{\rm Var}(\varphi)^{\rm Ind}_{\rm min}=\frac{(1+\alpha_{\psi})^{2}}{2\beta_{\psi}^{2}}=
\begin{cases}
0, & \text{if } \alpha_{\psi}=-1, \\[6pt]
2/\beta_{\psi}^{2}, & \text{if }  \alpha_{\psi}=1.
\end{cases}
\end{equation*}

\begin{table}[H]
\centering
\caption{The optimal values of the decay parameter $\alpha_{\psi}$ and scattering angle $\varphi$ correspond to the minimal variances for both simultaneous and individual estimations.}
\label{tab:optimal_variances}
\begin{center}
\begin{tabular}{c | c| c|| c |c |}
\cline{2-5}
 & ${\rm Var}(\varphi)_{\rm min}$ & ${\rm Var}(\varphi)^{\rm Ind}_{\rm min}$ & ${\rm Var}(\alpha_{\psi})_{\rm min}$ & ${\rm Var}(\alpha_{\psi})^{\rm Ind}_{\rm min}$ \\
\hline
$\varphi^{\rm opt}$ & $\pi/2$ &  $\pi/2$ &  $\pi/2$ &  $\pi/2$\\
\hline
$\alpha_{\psi}^{\rm opt}$ & $+1$ & $+1$ & $\pm 1$ & $\pm 1$\\
\hline
\end{tabular}
\end{center}
\end{table}

\begin{figure}[!h]
\includegraphics[scale=0.5]{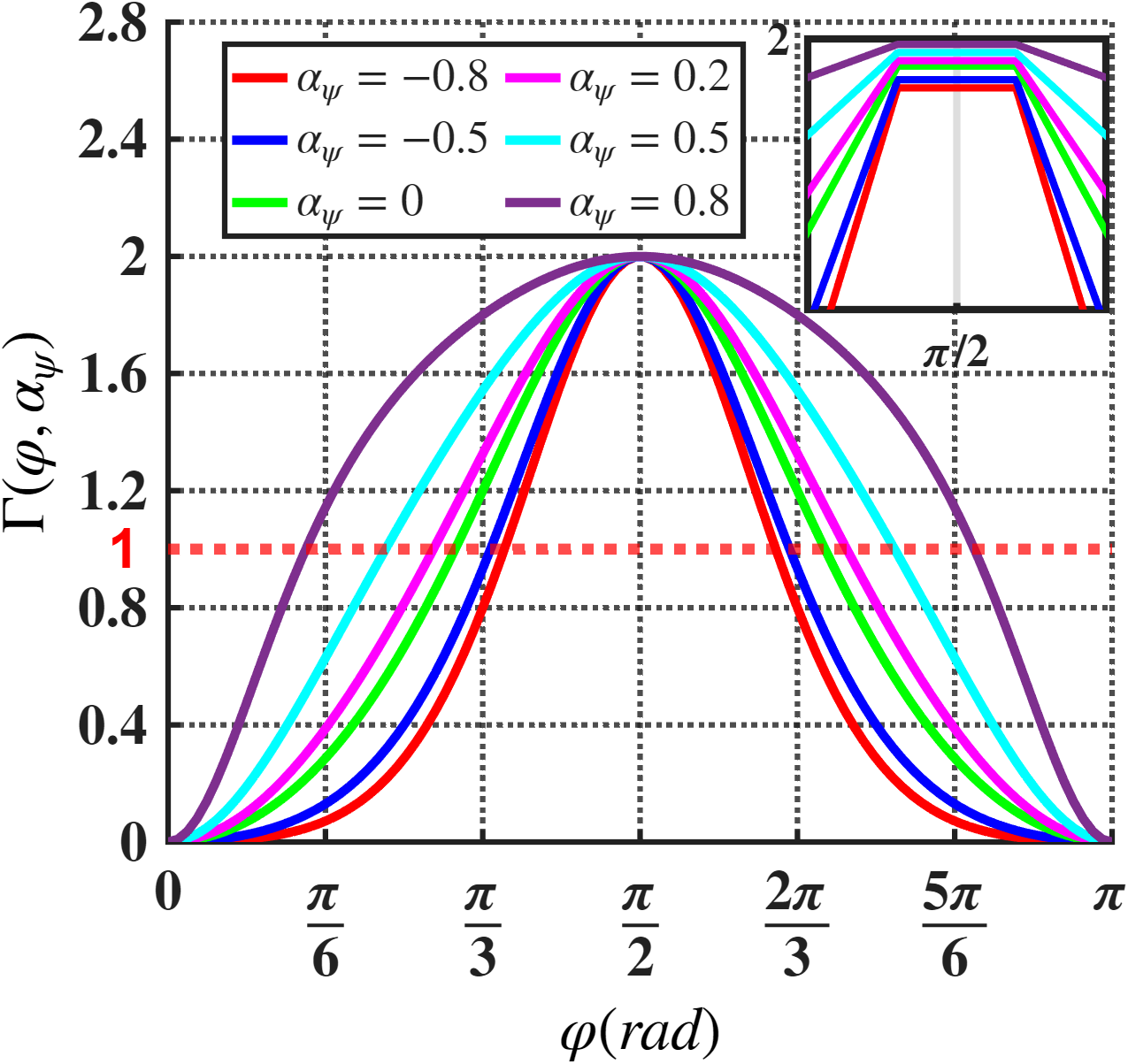}
\caption{Plot of the ratio $\Gamma(\varphi,\alpha_{\psi})$, in the high-energy limit ($\gamma_{\psi}=\beta_{\psi}=0$), between the minimal total variances for estimating $\varphi$ and $\alpha_{\psi}$.}
\label{fig:Lamda}
\end{figure}
The ratio of the total variances, used to compare the simultaneous and individual estimation approaches, is defined as
\begin{equation}
\Gamma(\varphi,\alpha_{\psi})  = \frac{{{\Delta _{\rm Sim}}}}{{{\Delta_{\rm Ind}}}}=\frac{{\rm{Var}}(\alpha_{\psi})_{\min }^{\rm Ind} + {\rm{Var}}(\varphi)_{\min }^{\rm Ind}}{\frac{1}{2}\left( {{\rm{Var}}{{( \alpha_{\psi})}_{\min }} + {\rm{Var}}{{(\varphi)}_{\min }}}\right)}, \label{Raport}
\end{equation}
If $\Gamma(\varphi, \alpha_{\psi}) > 1$ (implying $\Delta_{\text{Sim}} < \Delta_{\text{Ind}}$), simultaneous estimation yields higher precision than individual estimation. Conversely, if $\Gamma(\varphi, \alpha_{\psi}) < 1$ (implying $\Delta_{\text{Sim}} > \Delta_{\text{Ind}}$), the individual estimation approach is more effective.

In Fig.~\ref{fig:Lamda}, we explore the ratio $\Gamma(\varphi,\alpha_{\psi})$ versus the scattering angle $\varphi$ for various values of the decay parameter $\alpha_{\psi}$. It can be seen that $\Gamma(\varphi,\alpha_{\psi})$ is not always less than unity. Specifically, it exceeds one and reaches a value of 2 in the limit $\alpha_{\psi} \to 1$. The condition $\Gamma(\varphi,\alpha_{\psi}) \geq 1$ (i.e., $\text{Var}_{\rm Sim} \leq \text{Var}_{\rm Ind}$) indicates that the minimum variance achieved through simultaneous estimation is lower than the sum of the minimum variances from individually estimating $\varphi$ and $\alpha_{\psi}$. This clearly demonstrates that the simultaneous estimation strategy provides a significant precision advantage over estimating the parameters separately.

\subsection{Individual and Simultaneous Parameter Estimation in Correlated Quantum Channels}\label{sec:5} 

This section examines the dynamical evolution of the minimal variances for both simultaneous and individual estimation in the $e^{+}e^{-} \rightarrow \text{B}\bar{\text{B}}$ process. We analyze their dependence on the scattering angle $\varphi$, the decay parameter $\alpha_{\psi}$ of the $\psi(c\bar{c})$ charmonium, and the classical parameter $\mu$. As investigated by the BESIII experiment, the $e^{+}e^{-} \rightarrow J/\psi \rightarrow \text{B}\bar{\text{B}}$ channel produces ground-state octet baryon pairs; understanding the underlying parameters is crucial for characterizing their properties and interactions. We evaluate this estimation problem across Markovian and non-Markovian dynamical regimes. These regimes characterize the decoherence and relaxation processes governing the evolution of the $\text{B}\bar{\text{B}}$ system from production to decay.

The Markovian regime typically arises from weak coupling to noisy environments—such as stochastic scattering in detectors or magnetic-field fluctuations in collider experiments—and is commonly modeled by classical white noise \cite{M1, M2}. In contrast, the non-Markovian regime emerges in strongly coupled or structured environments, where time-correlated interactions can partially preserve quantum features \cite{M3}. By analyzing both regimes, we elucidate how experimental noise affects the ultimate precision limits of baryon decay parameter estimation, thereby establishing a link between open quantum system theory and collider phenomenology.

Starting with the initial state $\varrho^{\text{X}}_{\text{B}\bar{\text{B}}}(0)$, we assume each qubit evolves through the same channel $\Gamma$. The corresponding output state is determined by the map \cite{p4}
\begin{equation}
\hat{\varrho}^{\text{X}}_{\text{B}\bar{\text{B}}}(t) = \Gamma\big[\varrho^{\text{X}}_{\text{B}\bar{\text{B}}}(0)\big] = \sum_{i,j=0}^{3} \tilde{O}_{i,j} \varrho^{\text{X}}_{\text{B}\bar{\text{B}}}(0)\tilde{O}_{i,j}^{\dagger},
\label{eq:D}
\end{equation}
where The Kraus operators are written as $\tilde{O}_{i,j}=\sqrt{p_{i,j}} \tau_{i}\otimes\tau_{j}$. Here, $\tau_{0} = I_{2}$ is the identity matrix and $\tau_{x,y,z}$ correspond to the Pauli operators. The probabilities $p_{i}$ are non-negative and sum to one, thereby forming a valid probability distribution. As the map is CPTP, the Kraus operators $\tilde{O}_{i,j}$ obey the normalization condition $\sum_{i,j}\tilde{O}_{i,j}^\dagger \tilde{O}_{i,j}=\hat{\mathbb{I}}_{4}$, with the joint probability defined as
\begin{equation}
{\rm p}_{i,j}=(1-\mu){\rm p}_{i,j}+\mu {\rm p}_{i}\delta_{i,j},
\label{eq:p}
\end{equation}
where $\delta_{i,j}$ denotes the Kronecker delta, while $0 \leq \mu \leq 1$ represents the degree of classical correlation between successive applications of $\zeta$ to the qubits. To incorporate the effects of uncorrelated dephasing, we employ a channel with the following probability distribution: $p_{0}=1-p$, $p_{z}=p$, and $p_{x,y}=0$. We examine the time-dependent multipartite quantum estimation using a colored pure dephasing model characterized by a time-varying Hamiltonian \cite{p4}.
\begin{equation}
\hat{H}(t)=\hbar\mathcal{R}(t)\tau_{z},
\end{equation}
where the random telegraph signal $\mathcal{R}(t)=\omega \mathcal{D}(t)$ consists of a Poisson process $\mathcal{D}(t)$ with mean $\langle \mathcal{D}(t) \rangle = \frac{t}{2\tau}$ and a discrete random variable $\omega \in \{\pm\omega\}$. Taking $\omega=1$, the time-dependent parameter $p$ is expressed as
\begin{equation}
{\rm p}=\frac{1-\mathcal{G}(t)}{2},
\end{equation}
with  $v=\sqrt{|1-16\tau^2|}$. Moreover, for $\tau < 1/4$, which characterizes the Markovian regime, the expression becomes
\begin{equation}
\mathcal{G}(t)= \left\lbrace \cosh \frac{t}{2\tau}-\sinh\frac{t}{2\tau} \right\rbrace \left\lbrace \cosh vt +\frac{1}{v}\sinh vt\right\rbrace,
\label{eq:rt}
\end{equation}
in the non-Markovian case ($\tau > 1/4$), we obtain
\begin{equation}
\mathcal{G}(t)=\left\lbrace \cosh \frac{t}{2\tau}-\sinh\frac{t}{2\tau} \right\rbrace \left\lbrace \cos vt +\frac{1}{v}\sin vt\right\rbrace,
\label{eq:rtt}
\end{equation}
under the influence of correlated dephasing, the final state of the $\text{B}\bar{\text{B}}$ system is derived by applying the map in Eq.~(\ref{eq:D}) to the initial state Eq.~(\ref{eq:varrho}). Incorporating the probability defined in Eq.~(\ref{eq:p}), the density operator becomes
\begin{equation}
\begin{aligned}
\hat{\varrho}_{\text{B}\bar{\text{B}}}(t)=
\begin{bmatrix}
\varrho_{1,1}& 0 & 0 & \kappa\varrho_{1,4} \\
0 & \varrho_{2,2} &\kappa\varrho_{2,3} & 0 \\
0 & \kappa\varrho_{3,2} & \varrho_{3,3}& 0 \\
\kappa\varrho_{4,1} & 0 & 0 & \varrho_{4,4}
\end{bmatrix},
\label{eq:varrhot}
\end{aligned}
\end{equation}
where 
\begin{equation}
\kappa = \mathcal{G}^{2}(t)+\left[ 1-\mathcal{G}^{2}(t)\right]\mu.
\label{eq:kappa}
\end{equation}
To avoid repetition of the calculations in Section \ref{sec:2}, we note that the minimal variances in Eqs. (\ref{varind}) and (\ref{varind1}) are obtained simply by scaling the density matrix elements $\varrho_{1,4}$ and $\varrho_{2,3}$ by the factor $\kappa$.

\begin{figure}[!h]
\includegraphics[scale=0.34]{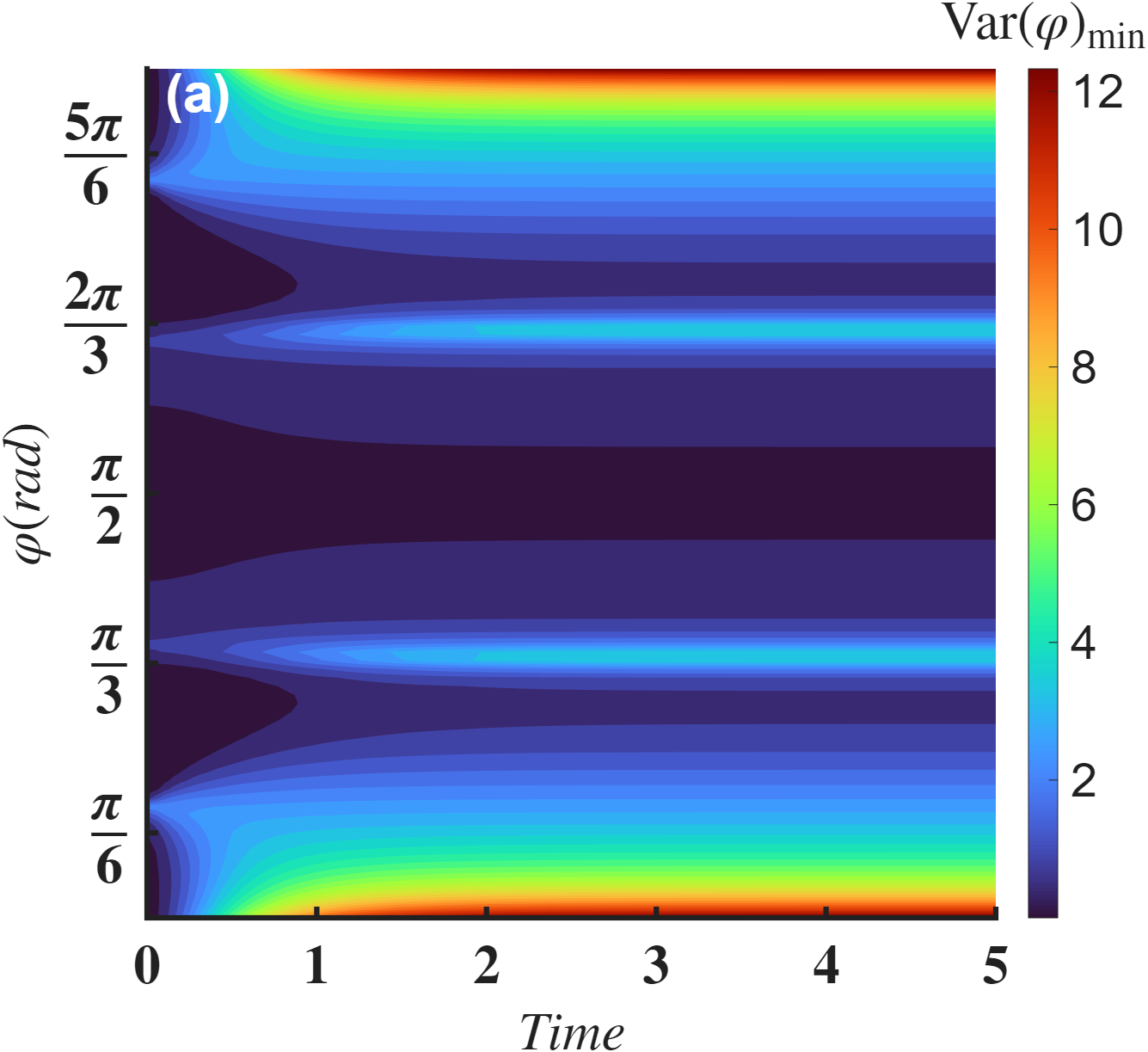}
\includegraphics[scale=0.34]{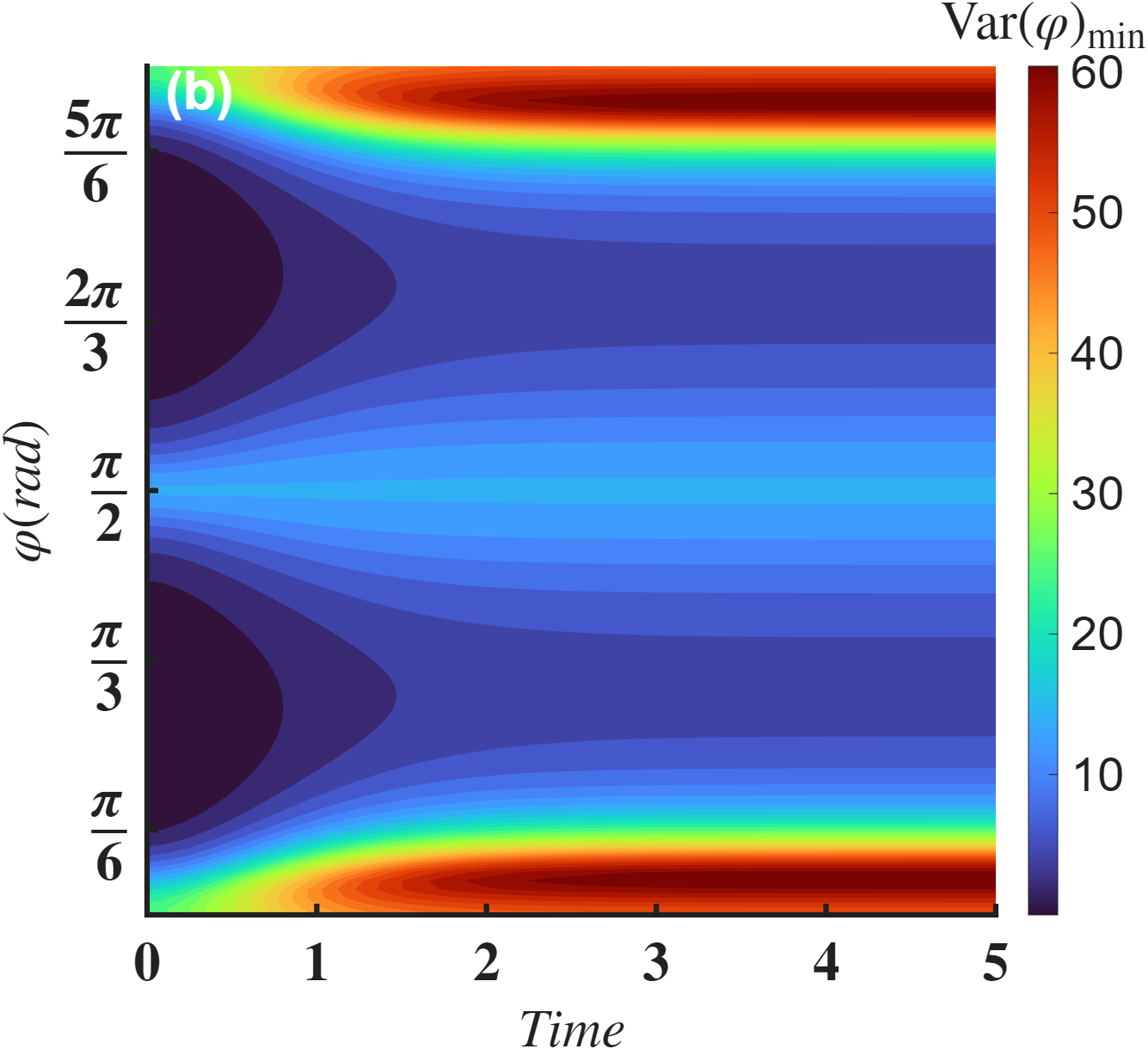}
\caption{Dynamical evolution the minimal variance $\mathrm{Var}(\varphi)_{\min}$ of the simultaneous estimates of $\varphi$ versus the scattering angle $\varphi$ in the Markovian regime ($\tau=0.2$ and $\mu=0.4$) for (a) $e^+e^-\to J/\psi\to \Xi^{0}\bar{\Xi}^{0}$, and (b) $e^+e^-\to J/\psi\to \Sigma^{+}\bar{\Sigma}^{-}$. The experimental parameter values are taken from Table \ref{tab1}.}
\label{fig:Mar1}
\end{figure}
\begin{figure}[!h]
\includegraphics[scale=0.34]{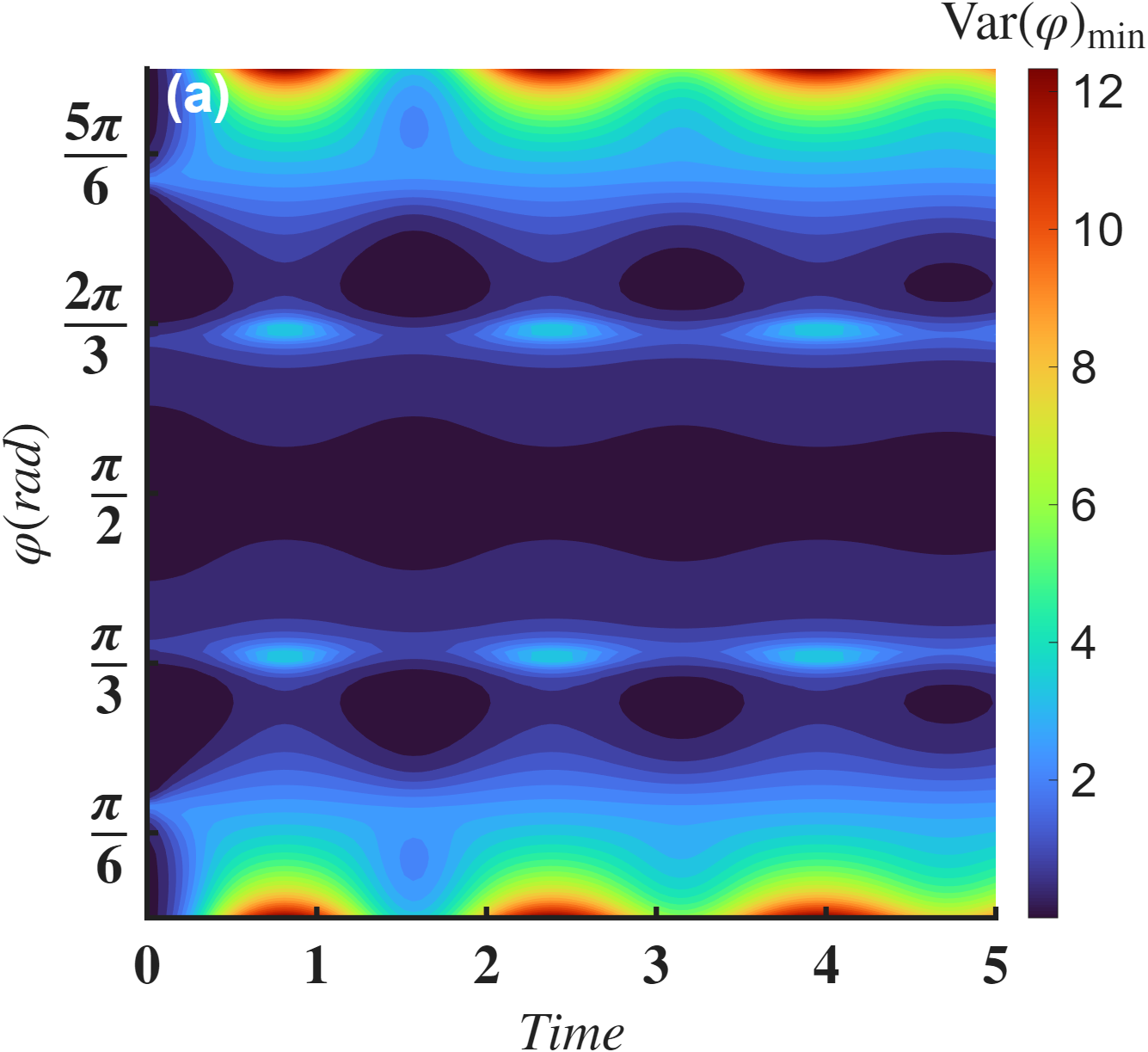}
\includegraphics[scale=0.34]{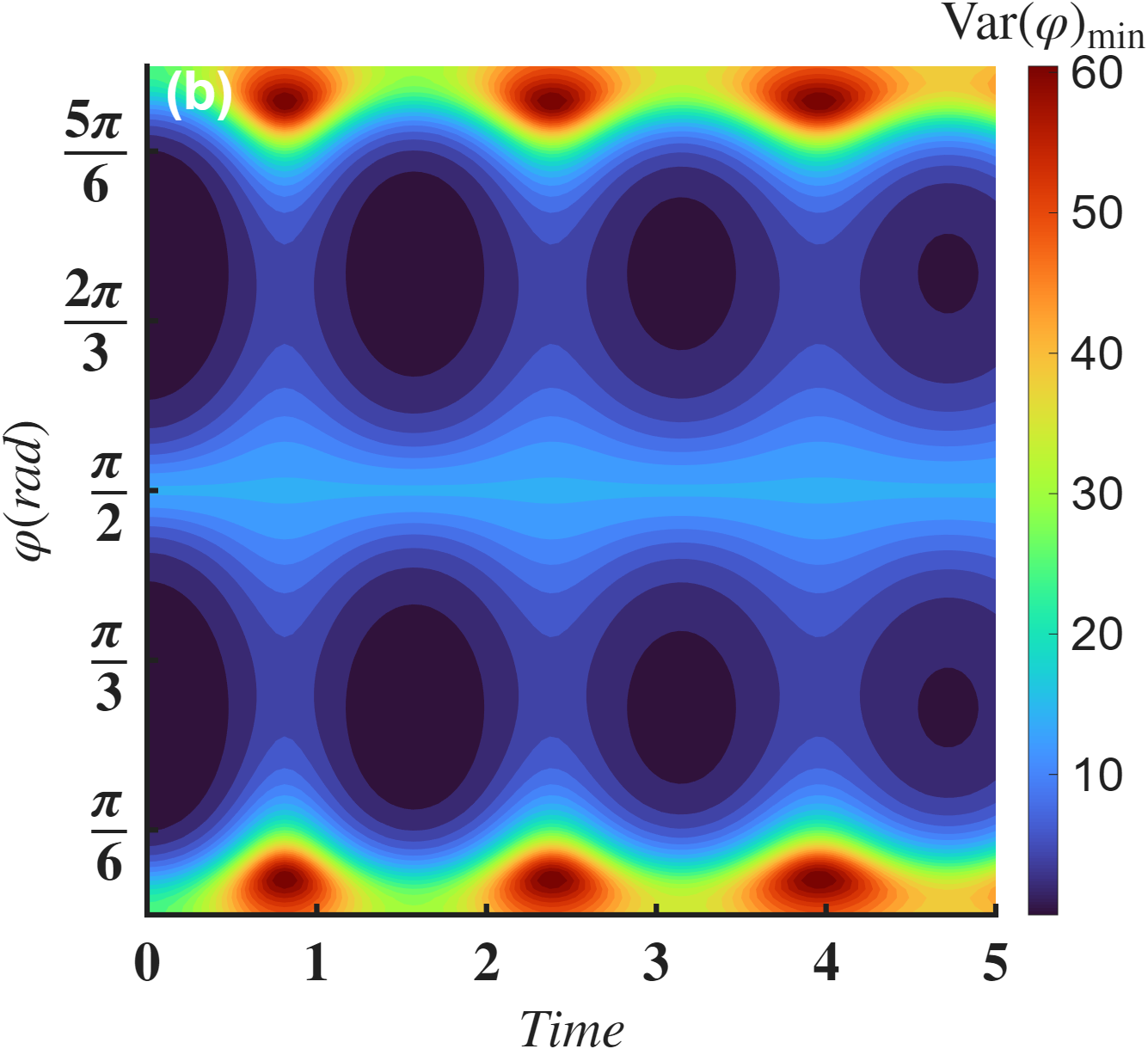}
\caption{Dynamical evolution of the minimal variance $\mathrm{Var}(\varphi)_{\min}$ for the simultaneous estimation of $\varphi$ versus the scattering angle $\varphi$ in the non-Markovian regime ($\tau=5$ and $\mu=0.4$) for (a) $e^+e^-\to J/\psi\to \Xi^{0}\bar{\Xi}^{0}$ and (b) $e^+e^-\to J/\psi\to \Sigma^{+}\bar{\Sigma}^{-}$. The parameter values are taken from Table \ref{tab1}.}
\label{fig:non-Mar1}
\end{figure}
In Figs.~\ref{fig:Mar1}(a--b), we illustrate the minimal variance $\mathrm{Var}(\varphi)_{\min}$ versus time and scattering angle $\varphi$ in the Markovian regime for the $e^{+}e^{-} \to \text{B}\bar{\text{B}}$ processes ($\text{B}=\Xi^{0}, \Sigma^{+}$) at BESIII. The variance peaks near $\varphi=0$ and $\pi$ due to the vanishing quantum Fisher information, while the highest precision (global minimum) is reached at $\varphi=\pi/2$. While the Markovian variance decays monotonically, the non-Markovian dynamics in Fig.~\ref{fig:non-Mar1} exhibit damped oscillations, reflecting environmental memory effects.

\begin{figure}[!h]
\includegraphics[scale=0.34]{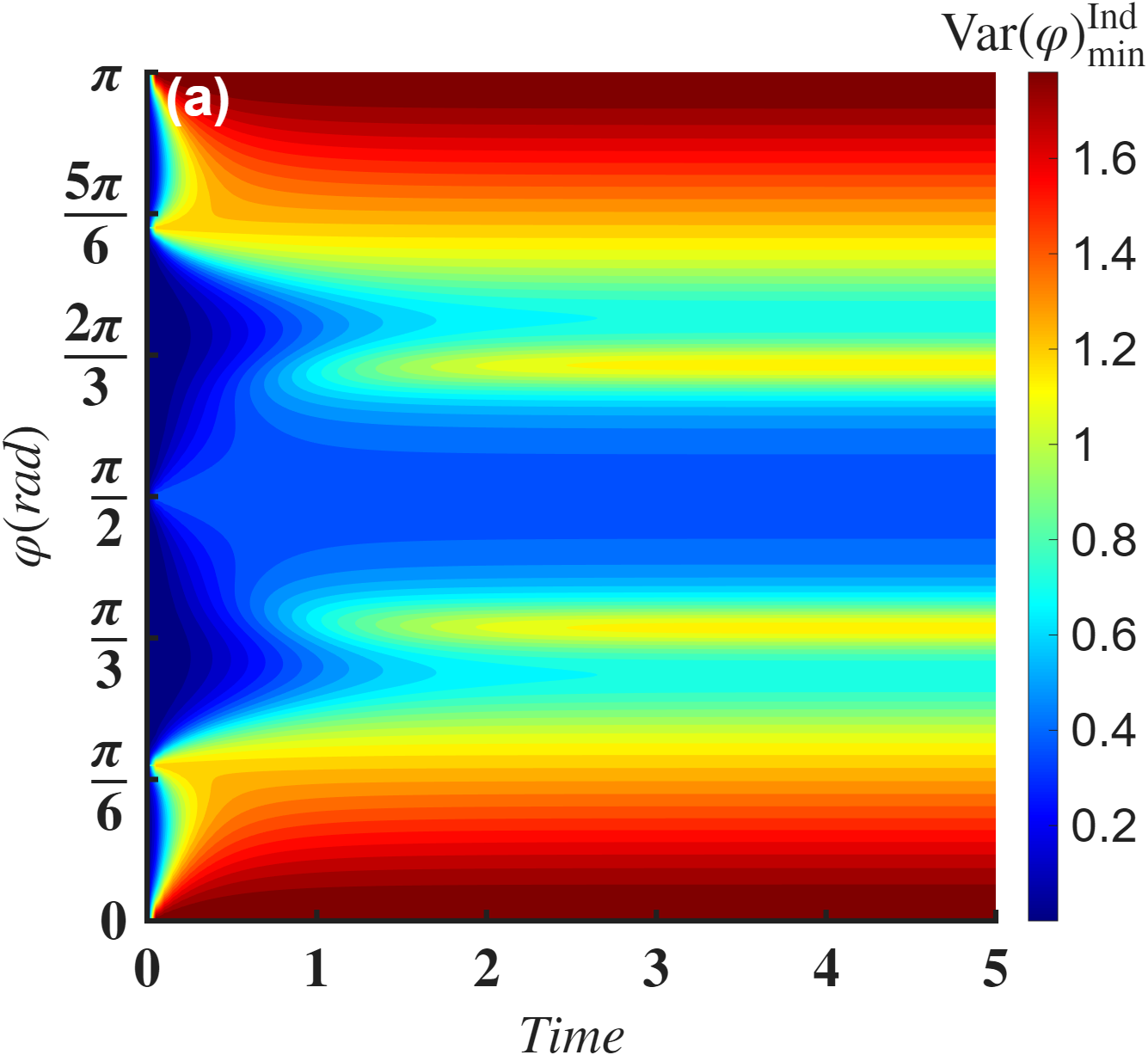}
\includegraphics[scale=0.34]{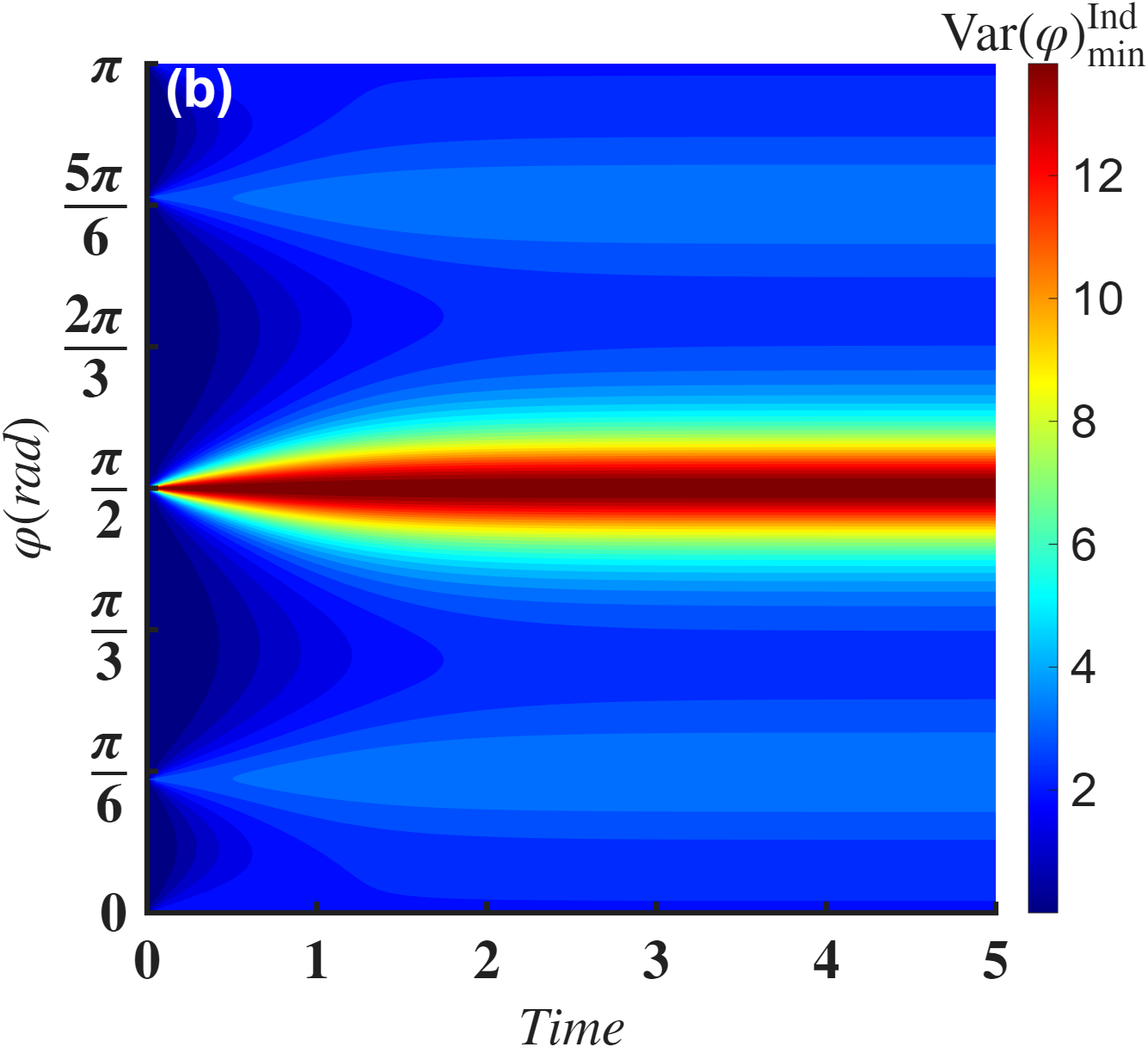}
\caption{Time evolution of the minimal variance for individual estimation as a function of the scattering angle $\varphi$ in the Markovian regime ($\tau=0.2$ and $\mu=0.4$) for (a) $e^+e^-\to J/\psi\to \Xi^{0}\bar{\Xi}^{0}$ and (b) $e^+e^-\to J/\psi\to \Sigma^{+}\bar{\Sigma}^{-}$. The parameter values are taken from Table \ref{tab1}.}
\label{fig:Mar2}
\end{figure}
\begin{figure}[!h]
\includegraphics[scale=0.34]{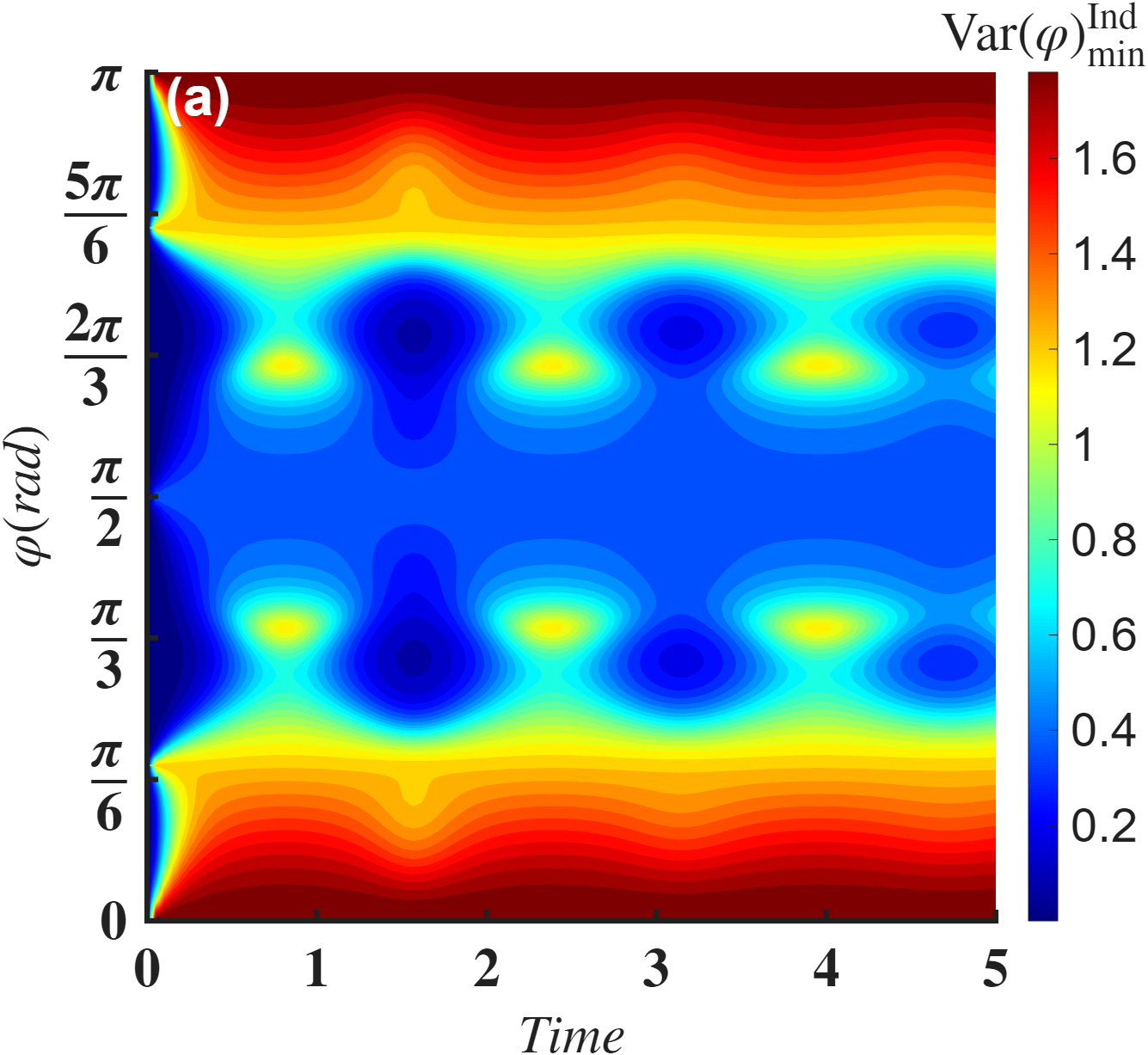}
\includegraphics[scale=0.34]{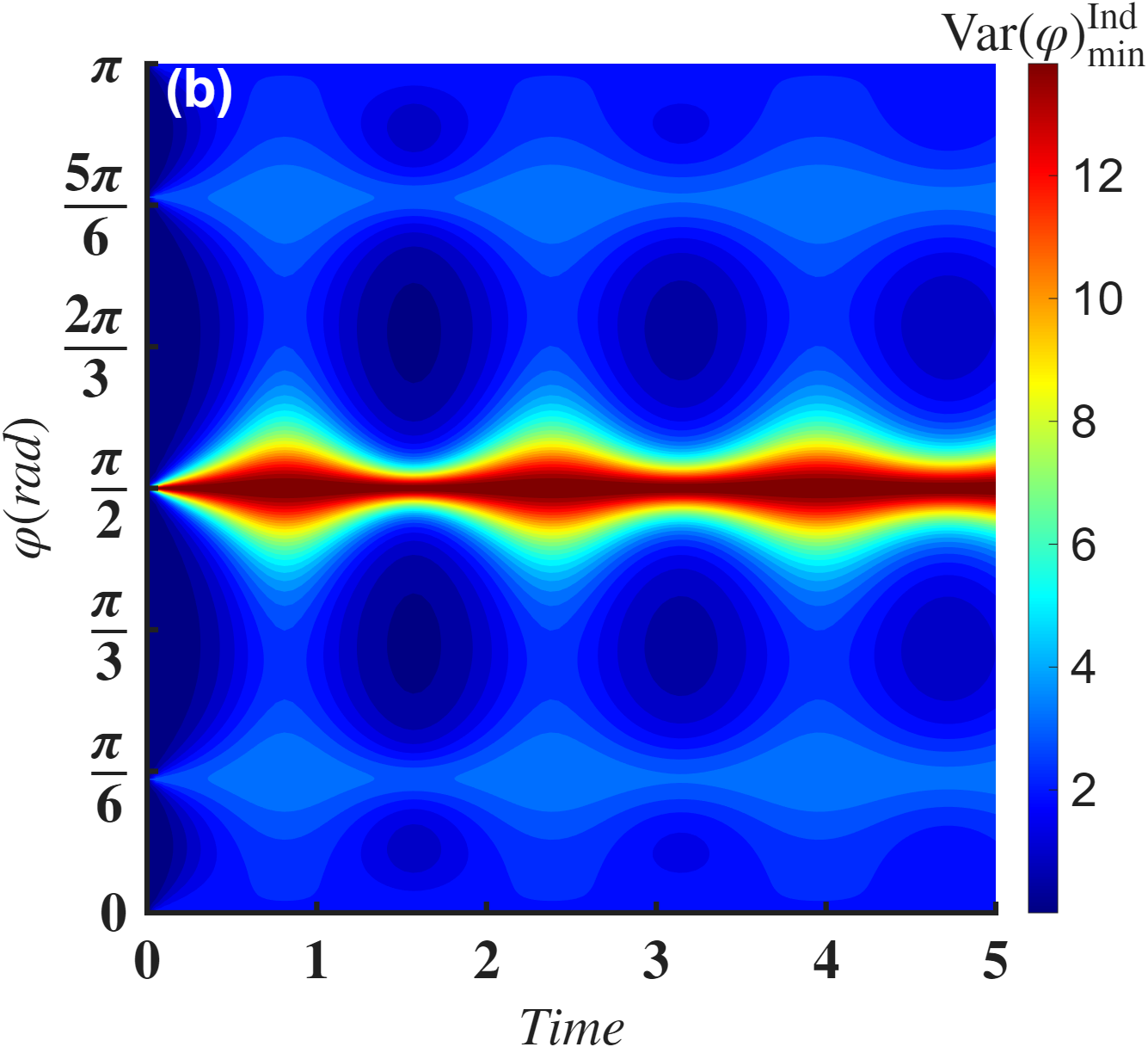}
\caption{Time evolution of the minimal variance for individual estimation as a function of the scattering angle $\varphi$ in the non-Markovian regime ($\tau=5$ and $\mu=0.4$) for (a) $e^+e^-\to J/\psi\to \Xi^{0}\bar{\Xi}^{0}$ and (b) $e^+e^-\to J/\psi\to \Sigma^{+}\bar{\Sigma}^{-}$. The parameter values are taken from Table \ref{tab1}.}
\label{fig:Non-Mar2}
\end{figure}
In Figs.~\ref{fig:Mar2}(a--b), we plot the minimal variance associated with the individual estimation of $\varphi$, denoted $\mathrm{Var}(\varphi)_{\min}^{\mathrm{Ind}}$, as a function of time and scattering angle $\varphi$ in the Markovian regime. As seen in Fig.~\ref{fig:Mar2}(a), the variance is minimized near $\varphi = \pi/2$, indicating high estimation precision. Conversely, the variance reaches its maximum around $\varphi = \{0, \pi\}$, where the quantum Fisher information for $\varphi$ is low. Interestingly, for $\Sigma^{+}\bar{\Sigma}^{-}$, the behavior is reversed: the maximum occurs at $\varphi = \pi/2$, as shown in Fig.~\ref{fig:Mar2}(b). Moreover, the Markovian variance decays monotonically, the non-Markovian dynamics in Fig.~\ref{fig:Non-Mar2} exhibit damped oscillations, reflecting environmental memory effects. Additionally, at $\varphi=\pi/2$, the terms $F_{\varphi\alpha_{\psi}}$ and $F_{\alpha_{\psi}\varphi}$ disappear, resulting in a diagonal quantum Fisher information matrix (QFIM), as given by
\begin{equation}
F=
\begin{pmatrix}
F_{\alpha_{\psi}\alpha_{\psi}} & 0\\
0 & F_{\varphi\varphi}
\end{pmatrix}.
\end{equation}
This implies
\begin{subequations}
\begin{align}
\mathrm{Var}(\varphi)_{\min} = \mathrm{Var}(\varphi)_{\min}^{\mathrm{Ind}} = \frac{1}{F_{\varphi\varphi}}, 
\end{align}
and
\begin{align}
\quad \mathrm{Var}(\alpha_{\psi})_{\min} = \mathrm{Var}(\alpha_{\psi})_{\min}^{\mathrm{Ind}} = \frac{1}{F_{\alpha_{\psi}\alpha_{\psi}}}.
\end{align}
\end{subequations}
The simultaneous and individual Cram\'er-Rao bounds both attain their minimum at $\varphi = \pi/2$, signifying that the estimation of $\varphi$ decouples from the remaining parameters, as expected from the diagonal form of the quantum Fisher information matrix at this angle. 
\begin{figure}[!h]
\includegraphics[scale=0.38]{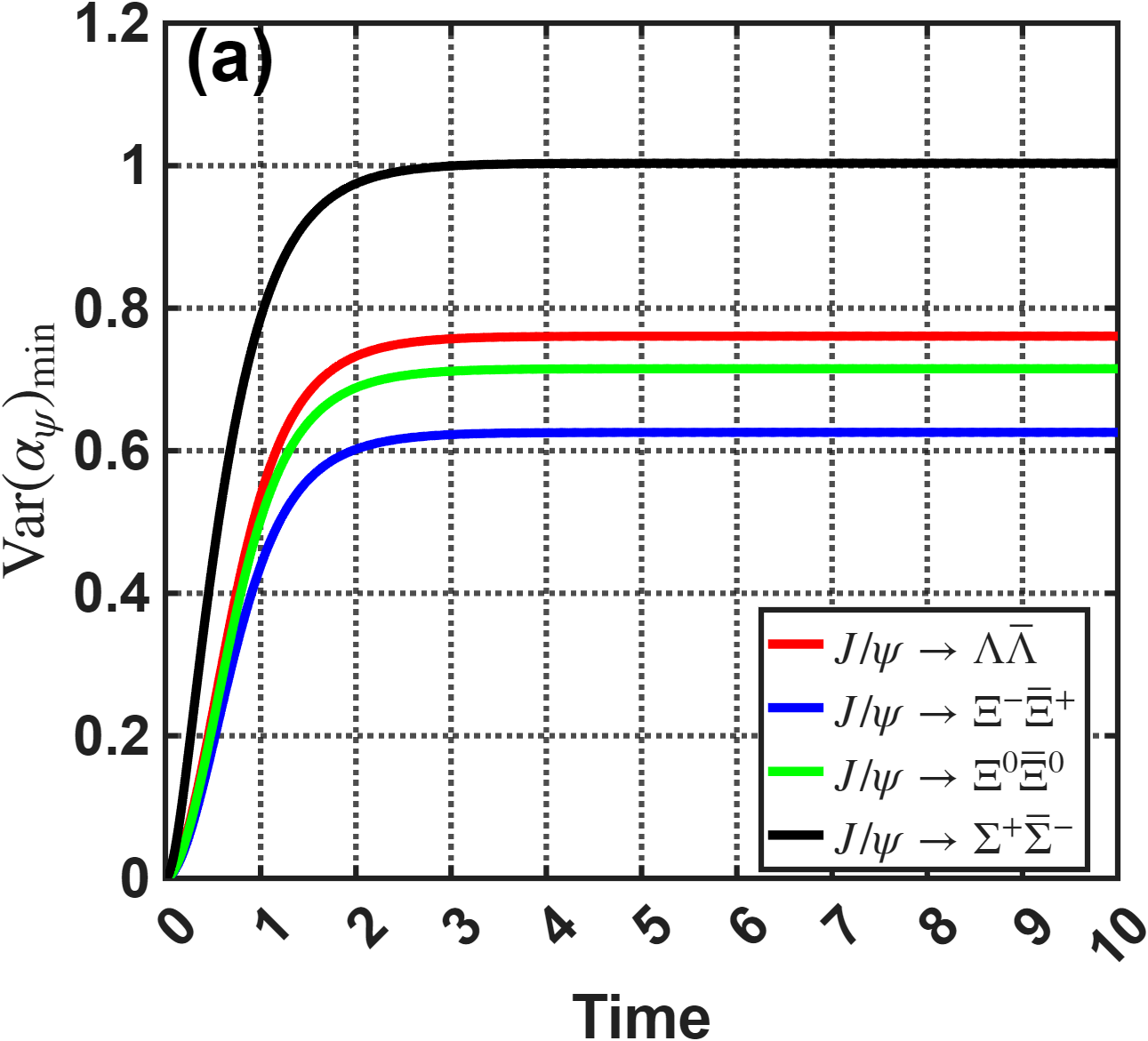}
\includegraphics[scale=0.38]{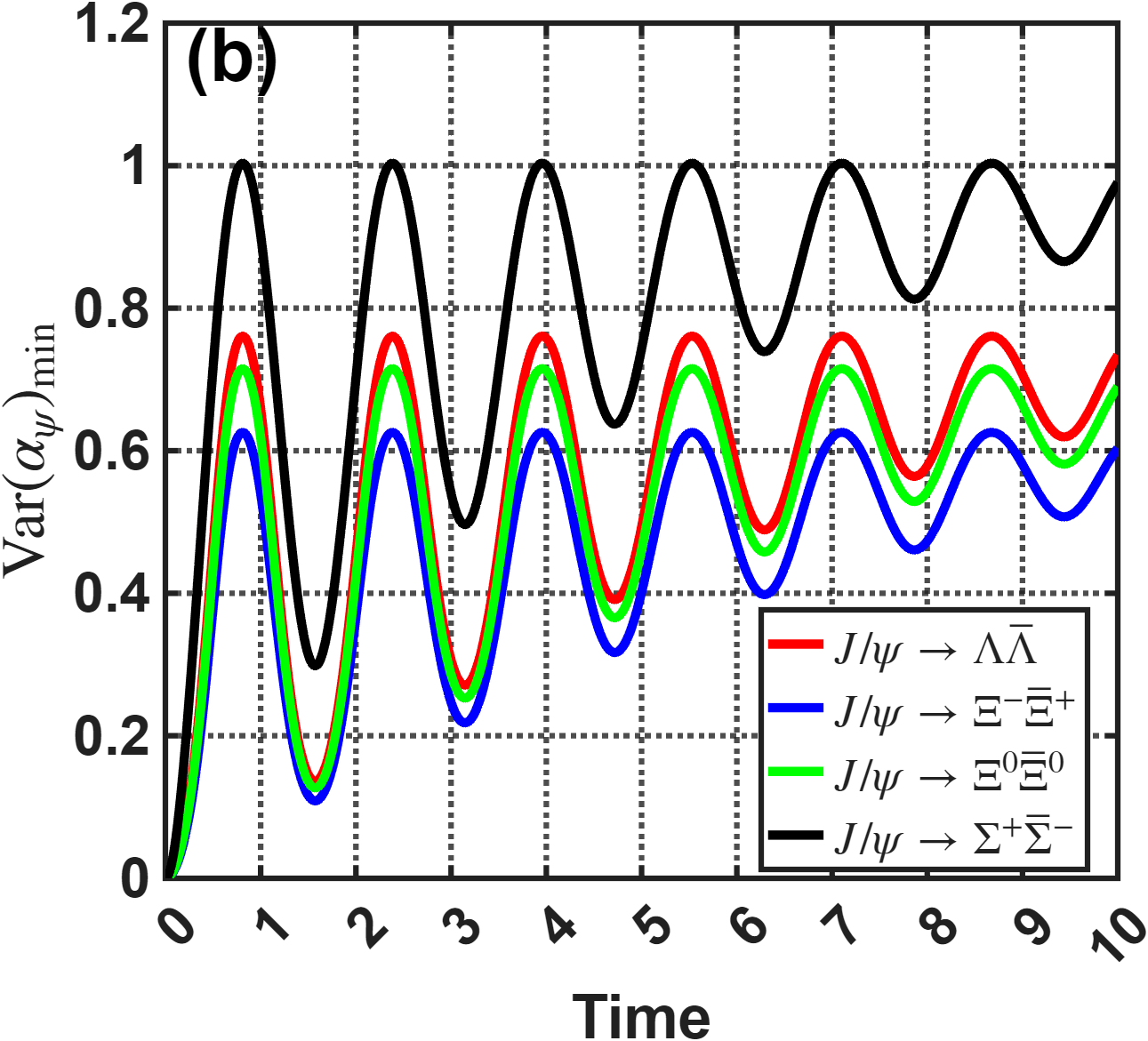}
\caption{Time evolution of the simultaneous estimates for the decay parameter $\alpha_{\psi}$, with $\mu = 0.2$ and $\varphi=\pi/2$, in (a) the Markovian regime ($\tau= 0.2$) and (b) the non-Markovian regime ($\tau= 5$) for the process $e^+e^-\to J/\psi\to \text{B}\bar{\text{B}}$. The parameter values are taken from Table \ref{tab1}.}
\label{fig:Mar3}
\end{figure}

The dynamical evolution of the minimal variance for the $\text{B}\bar{\text{B}}$ system is shown in Figs.~\ref{fig:Mar3}(a) and (b) for Markovian and non-Markovian noise, respectively. In the Markovian case [Fig.~\ref{fig:Mar3}(a)], $\mathrm{Var}(\alpha_{\psi})_{\min}$ increases monotonically toward a steady state. The high precision near $t=0$ ($\mathrm{Var}(\alpha_{\psi})_{\min} \to 0$) stems from the strong initial $\text{B}\bar{\text{B}}$ entanglement. Conversely, Fig.~\ref{fig:Mar3}(b) shows that non-Markovian dynamics induce damped oscillations in the variance due to memory effects. In both regimes, the highest precision corresponds to the limit where the variance approaches zero.

\section{Conclusion}\label{sec:6}

In summary, we have investigated multiparameter quantum estimation in the $e^+e^- \to J/\psi \to \text{B}\bar{\text{B}}$ process using the SLD operator formalism. A systematic comparison between individual and simultaneous estimation strategies was conducted to establish their relative performance. Our results demonstrate that estimation variances are minimized at a scattering angle of $\varphi = \pi/2$ and a decay parameter of $\alpha_\psi = 1$. The introduction of experimentally feasible classical correlations notably suppresses these variances. Furthermore, non-Markovian memory effects prove crucial for maintaining the stability of the estimation performance. Finally, we observed a general temporal growth in the variances at parameter-dependent rates, reflecting the progressive diffusion of information during the system's evolution.

\section*{Appendix V} \label{AppA}

\begin{equation*}
\begin{aligned}
\bar{\Lambda}_{1,1} &= 
\frac{-\varrho_{14}^{2} + \varrho_{44}^{2} + \varrho_{11}\varrho_{44}}{2(\varrho_{11}+\varrho_{44})(\varrho_{11}\varrho_{44}-\varrho_{14}^{2})}, \quad \bar{\Lambda}_{4,4} = \bar{\Lambda}_{13,13} = \frac{2\varrho_{11}\varrho_{44} - \varrho_{14}^{2}}{2(\varrho_{11}+\varrho_{44})(\varrho_{11}\varrho_{44}-\varrho_{14}^{2})}, \\[2mm]
\bar{\Lambda}_{16,16} &= 
\frac{\varrho_{11}^{2} + \varrho_{44}\varrho_{11} - \varrho_{14}^{2}
}{2(\varrho_{11}+\varrho_{44})(\varrho_{11}\varrho_{44}-\varrho_{14}^{2})}, \quad \bar{\Lambda}_{1,4} = \bar{\Lambda}_{1,13} = \bar{\Lambda}_{4,1} = \bar{\Lambda}_{13,1} = 
-\frac{\varrho_{14}\varrho_{44}}{2(\varrho_{11}+\varrho_{44})(\varrho_{11}\varrho_{44}-\varrho_{14}^{2})}, \\[2mm]
\bar{\Lambda}_{1,16} &= \bar{\Lambda}_{4,13} =
\bar{\Lambda}_{13,4} = \bar{\Lambda}_{16,1} =
\frac{\varrho_{14}^{2}}{2(\varrho_{11}+\varrho_{44})(\varrho_{11}\varrho_{44}-\varrho_{14}^{2})}, \quad \bar{\Lambda}_{4,16} = \bar{\Lambda}_{16,4} =\bar{\Lambda}_{13,16} = \bar{\Lambda}_{16,13} =-\frac{\varrho_{11}\varrho_{14}}{2(\varrho_{11}+\varrho_{44})(\varrho_{11}\varrho_{44}-\varrho_{14}^{2})}, \\[4mm]
\bar{\Lambda}_{2,2} &= \bar{\Lambda}_{3,3} =
\bar{\Lambda}_{5,5} = \bar{\Lambda}_{9,9} =
\frac{-\varrho_{14}^{2}\varrho_{22} - \varrho_{14}^{2}\varrho_{44} + 2\varrho_{22}^{2}\varrho_{44} + \varrho_{22}\varrho_{44}^{2}
 + 2\varrho_{11}\varrho_{22}\varrho_{44} + \varrho_{11}\varrho_{44}^{2}}{(\varrho_{11}\varrho_{44}-\varrho_{14}^{2})
 \left(2\varrho_{11}\varrho_{22} + \varrho_{11}\varrho_{44}
 + 2\varrho_{22}\varrho_{44} - \varrho_{14}^{2}
 + 4\varrho_{22}^{2}\right)}, \\[4mm]
\bar{\Lambda}_{2,3} &= \bar{\Lambda}_{3,2} =
\bar{\Lambda}_{5,9} = \bar{\Lambda}_{9,5} =
-\frac{\varrho_{22}\left(\varrho_{14}^{2} + \varrho_{44}^{2}
 + 2\varrho_{22}\varrho_{44}\right)}{(\varrho_{11}\varrho_{44}-\varrho_{14}^{2})\left(2\varrho_{11}\varrho_{22} + \varrho_{11}\varrho_{44}+ 2\varrho_{22}\varrho_{44} - \varrho_{14}^{2}+ 4\varrho_{22}^{2}\right)},\\[4mm] 
\bar{\Lambda}_{2,14}&=\bar{\Lambda}_{3,15}=
\bar{\Lambda}_{5,8}=\bar{\Lambda}_{9,12}=
\bar{\Lambda}_{14,2}=\bar{\Lambda}_{15,3} =
-\frac{\varrho_{14}\left(\varrho_{11}\varrho_{22} + \varrho_{11}\varrho_{44} + \varrho_{22}\varrho_{44}-\varrho_{14}^{2} + 2\varrho_{22}^{2}\right)}{(\varrho_{11}\varrho_{44}-\varrho_{14}^{2})\left(2\varrho_{11}\varrho_{22} + \varrho_{11}\varrho_{44}+ 2\varrho_{22}\varrho_{44} - \varrho_{14}^{2} + 4\varrho_{22}^{2}\right)}, \\[4mm]
\bar{\Lambda}_{2,15}&=\bar{\Lambda}_{3,14}=
\bar{\Lambda}_{5,12}=\bar{\Lambda}_{9,8}=
\bar{\Lambda}_{14,3}=\bar{\Lambda}_{15,2} =
\frac{\varrho_{14}\varrho_{22}(\varrho_{11}+2\varrho_{22}+\varrho_{44})
}{(\varrho_{11}\varrho_{44}-\varrho_{14}^{2})
 \left(2\varrho_{11}\varrho_{22} + \varrho_{11}\varrho_{44}
 + 2\varrho_{22}\varrho_{44} - \varrho_{14}^{2}
 + 4\varrho_{22}^{2}\right)}, \\[4mm]
\bar{\Lambda}_{8,8}&=\bar{\Lambda}_{12,12}=
\bar{\Lambda}_{14,14}=\bar{\Lambda}_{15,15} =
\frac{\varrho_{11}^{2}\varrho_{22} + \varrho_{44}\varrho_{11}^{2}
 -\varrho_{11}\varrho_{14}^{2} + 2\varrho_{11}\varrho_{22}^{2}
 + 2\varrho_{44}\varrho_{11}\varrho_{22}
 -\varrho_{14}^{2}\varrho_{22}}{(\varrho_{11}\varrho_{44}-\varrho_{14}^{2})\left(2\varrho_{11}\varrho_{22} + \varrho_{11}\varrho_{44}+ 2\varrho_{22}\varrho_{44} - \varrho_{14}^{2}+ 4\varrho_{22}^{2}\right)}, \\[4mm]
\bar{\Lambda}_{8,9}&=\bar{\Lambda}_{12,5}=
\bar{\Lambda}_{14,15}=\bar{\Lambda}_{15,14} =
\frac{\varrho_{14}\varrho_{22}(\varrho_{11}+2\varrho_{22}+\varrho_{44})
}{(\varrho_{11}\varrho_{44}-\varrho_{14}^{2})
 \left(2\varrho_{11}\varrho_{22} + \varrho_{11}\varrho_{44}
 + 2\varrho_{22}\varrho_{44} - \varrho_{14}^{2}
 + 4\varrho_{22}^{2}\right)}, \\[4mm]
\bar{\Lambda}_{8,12}&=\bar{\Lambda}_{12,8}=
\bar{\Lambda}_{14,15}=\bar{\Lambda}_{15,14} =
-\frac{\varrho_{22}(\varrho_{11}^{2} + 2\varrho_{22}\varrho_{11} + \varrho_{14}^{2})}{(\varrho_{11}\varrho_{44}-\varrho_{14}^{2})
 \left(2\varrho_{11}\varrho_{22} + \varrho_{11}\varrho_{44}
 + 2\varrho_{22}\varrho_{44} - \varrho_{14}^{2}
 + 4\varrho_{22}^{2}\right)
}, \\[4mm]
\bar{\Lambda}_{6,6} &= \bar{\Lambda}_{7,7} =
\bar{\Lambda}_{10,10} = \bar{\Lambda}_{11,11}
= \frac{5}{16\varrho_{22}}, \quad \bar{\Lambda}_{6,11} = \bar{\Lambda}_{7,10} =
\bar{\Lambda}_{11,6} = \bar{\Lambda}_{10,7}
= -\frac{3}{16\varrho_{22}},\\[3mm]
\bar{\Lambda}_{6,7} &= \bar{\Lambda}_{6,10} =
\bar{\Lambda}_{7,6} = \bar{\Lambda}_{7,11} =
\bar{\Lambda}_{10,6} = \bar{\Lambda}_{10,11} =
\bar{\Lambda}_{11,7} = \bar{\Lambda}_{11,10}
= \frac{1}{16\varrho_{22}}.
\end{aligned}
\end{equation*}

\newpage
\bibliography{sample}


\end{document}